\documentclass[fleqn,usenatbib]{mnras}

\usepackage{newtxtext,newtxmath}

\usepackage[T1]{fontenc}
\usepackage{ae,aecompl}

\usepackage{graphicx}	
\usepackage{amsmath}	
\usepackage{amssymb}	
\usepackage{float}

\title[mid-IR to sub-mm variability of YSOs]{The Relationship between Mid-Infrared and Sub-Millimetre Variability of Deeply Embedded Protostars}

\author[Contreras Pe\~{n}a et al.]{
Carlos Contreras Pe\~{n}a,$^{1}$\thanks{E-mail: c.contreras@exeter.ac.uk (CCP)}
Doug Johnstone,$^{2,3}$
Giseon Baek,$^{4}$
Gregory J. Herczeg,$^{5}$
\newauthor Steve Mairs,$^{6}$
Aleks Scholz,$^{7}$ 
Jeong-Eun Lee$^{4}$ and The JCMT Transient Team
\\
$^{1}$School of Physics, Astrophysics Group, University of Exeter, Stocker Road, Exeter EX4 4QL, UK\\
$^{2}$NRC Herzberg Astronomy and Astrophysics, 5071 West Saanich Road, Victoria, BC, V9E 2E7, Canada\\
$^{3}$Department of Physics and Astronomy, University of Victoria, Victoria, BC, V8P 5C2, Canada\\
$^{4}$School of Space Research and Institute of Natural Sciences, Kyung Hee University, 1732 Deogyeong-daero, Giheung-gu, Yongin-si, Gyeonggi-do 446-701, Korea\\
$^{5}$Kavli Institute for Astronomy and Astrophysics, Peking University, Yiheyuan 5, Haidian Qu, 100871 Beijing, People's Republic of China\\
$^{6}$East Asian Observatory, 660 North Aohoku Place, University Park, Hilo, HI 96720, USA\\
$^{7}$SUPA, School of Physics \& Astronomy, North Haugh, St Andrews KY16 9SS, UK
}

\date{Accepted XXX. Received YYY; in original form ZZZ}

\pubyear{2020}

\begin{document}
\label{firstpage}
\pagerange{\pageref{firstpage}--\pageref{lastpage}}
\maketitle

\begin{abstract}
We study the relationship between the mid-infrared and sub-mm variability of deeply embedded protostars using the multi-epoch data from the Wide Infrared Survey Explorer ({\it WISE}/NEOWISE) and the ongoing James Clerk Maxwell Telescope (JCMT) transient survey. Our search for signs of stochastic (random) and/or secular (roughly monotonic in time) variability in a sample of 59 young stellar objects (YSOs) revealed that 35 are variable in at least one of the two surveys. This variability is dominated by secular changes.  Of those objects with secular variability, 14 objects ($22\%$ of the sample) show correlated secular variability over mid-IR and sub-mm wavelengths. Variable accretion is the likely mechanism responsible for this type of variability. Fluxes of YSOs that vary in both wavelengths follow a relation of $\log_{10} F_{4.6}(t)=\eta \log_{10} F_{850}(t)$ between the mid-IR and sub-mm, with $\eta=5.53\pm0.29$. This relationship arises from the fact that sub-mm fluxes respond to the dust temperature in the larger envelope whereas the mid-IR emissivity is more directly proportional to the accretion luminosity.  The exact scaling relation, however, depends on the structure of the envelope, the importance of viscous heating in the disc, and dust opacity laws.
\end{abstract}




\section{Introduction}

Variable accretion in young stellar objects (YSOs) is produced by different physical mechanisms, each with outburst events with a range of amplitudes and timescales. Instabilities in the magnetospheric accretion at the star-disc interface lead to stochastic accretion outbursts where flux increases by a factor of  5-10 over timescales of days \citep{2008Romanova, 2014Stauffer}, as seen in high-cadence monitoring \citep[e.g.][]{2017Cody}. Outbursts with duration from weeks to 100 years, thought to be driven by disc instabilities, are observed in the class of eruptive YSOs \citep[the EXors, MNors, and FUors, e.g.][]{1996Hartmann,2014Audard,2017Contreras}. 

Accretion variability in YSOs, especially the most extreme events (the FUors), has been suggested to explain many long-standing problems in the formation and evolution of these systems. If YSOs spend most of their lifetime in quiescent states and gain most of their mass in short-lived high accretion states, then this variability could solve the so-called ``luminosity problem'' observed in Class I YSOs \citep{1990Kenyon, 2009Evans} and explain the scatter observed around the best fitting isochrone in pre-MS clusters \citep{2012Baraffe,2017Baraffe,2017Kunimoto}. In addition, the luminosity bursts can impact the formation of brown dwarfs \citep{2012Stamatellos}, the formation and evolution of protoplanetary systems \citep{2016Cieza,2017Hubbardb}, as well as have an effect on the chemistry of protoplanetary discs in YSOs \citep{2007JLee, 2012Kim,2015Harsono, 2019Artur}.

Observational estimates of the rate of rare FUor outbursts (and consequently of outbursts with timescales of days to up to 10 years) suggest that accretion-driven outbursts are more frequent during the earlier stages of young stellar evolution (every $\sim10000$ years in the Class I stage) than during later stages \citep[every $\sim100000$ years during the Class II stage, see][]{2013Scholz,2019Fischer,2019Contreras}. Observations of knots along outflows of YSOs, which can be considered as fossil records of accretion outbursts during the most embedded phase (Class 0), suggest that these occur on timescales of $\sim$ 1000 years \citep[]{2018Makin}. The higher frequency of outbursts at earlier stages of YSO evolution agree with the expectation from theoretical models of transport of angular momentum in accretion discs \citep[see e.g. the gravitational instability models of][]{2014Bae, 2015Vorobyov}.

However, YSOs at the early stages are still deeply embedded in their nascent, dusty envelopes and are thus too heavily extincted for an outburst to be directly measured at optical or near-IR wavelengths. In these deeply embedded objects the radiation at shorter wavelengths is absorbed by the dense envelope and re-radiated at longer wavelengths, with the strongest signal arising at the mid- and far-IR and the sub-millimetre. The increase in flux at the longer wavelengths traces the heating of the envelope due to the accretion-driven outburst (e.g \citealt{2013Johnstone}; \citealt{2019Macfarlane_a,2019Macfarlane_b,2020Baek}).

\begin{figure*}
	\resizebox{1\columnwidth}{!}{\includegraphics{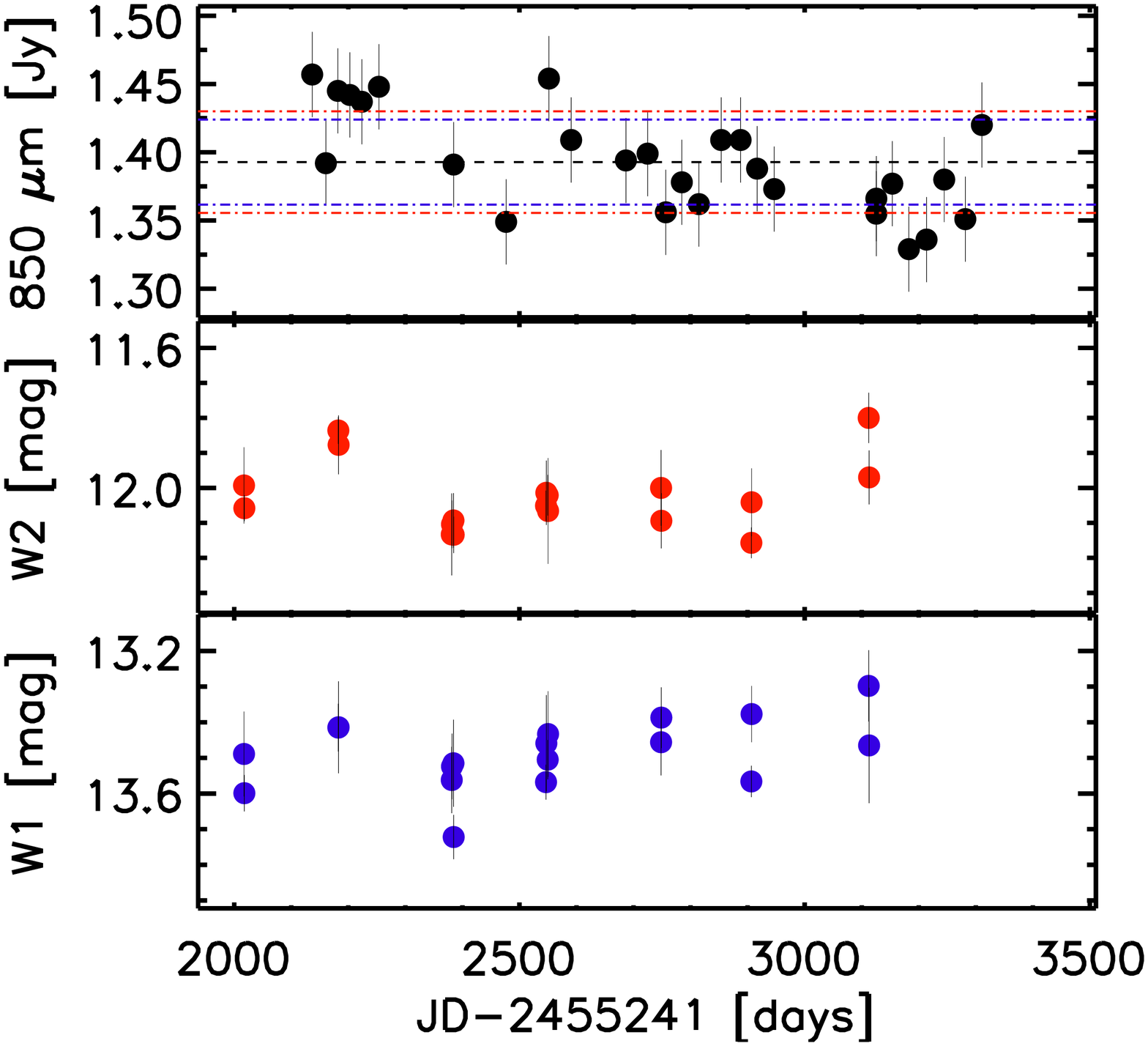}}
	\resizebox{0.95\columnwidth}{!}{\includegraphics{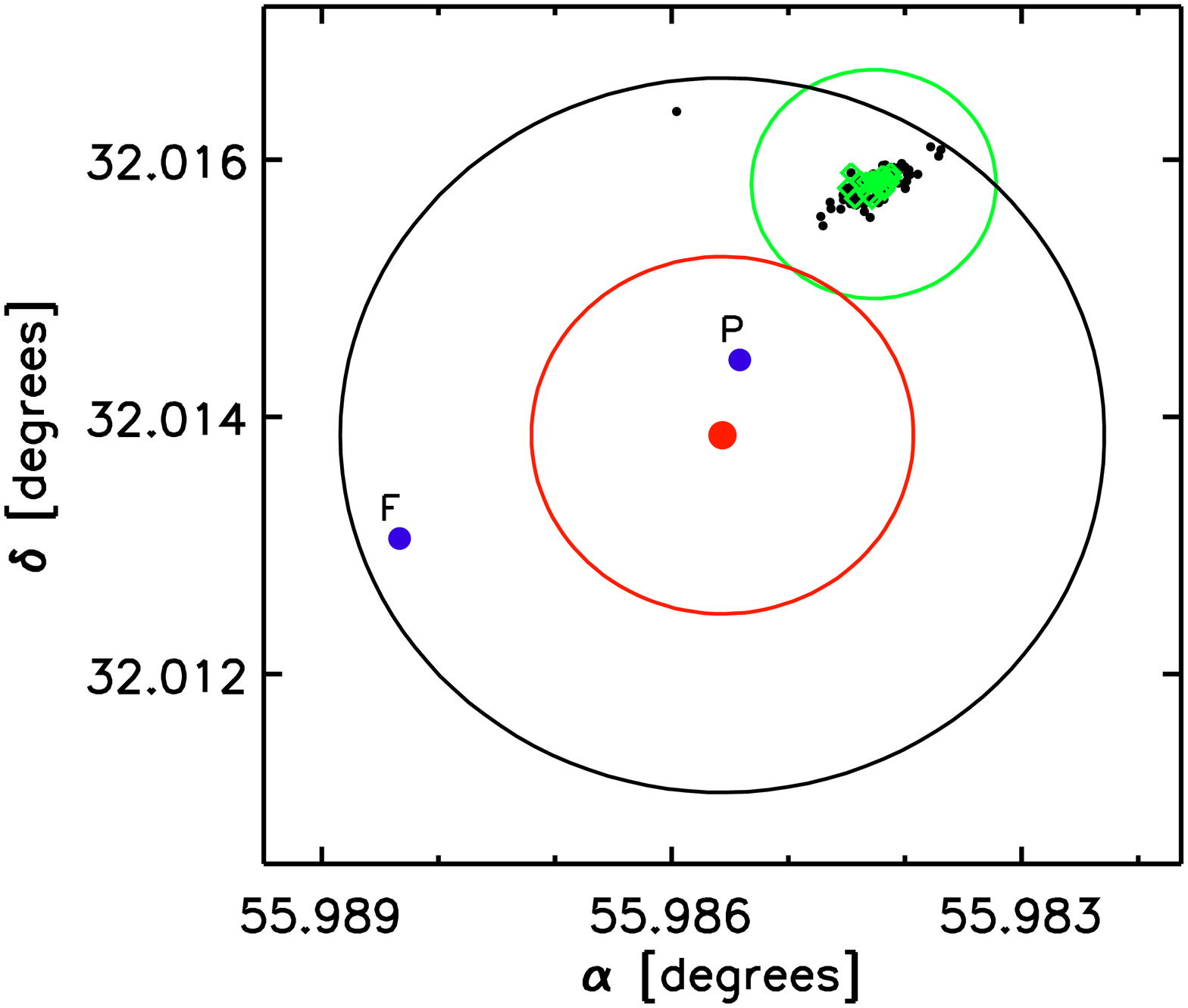}} \\
	\resizebox{2\columnwidth}{!}{\includegraphics{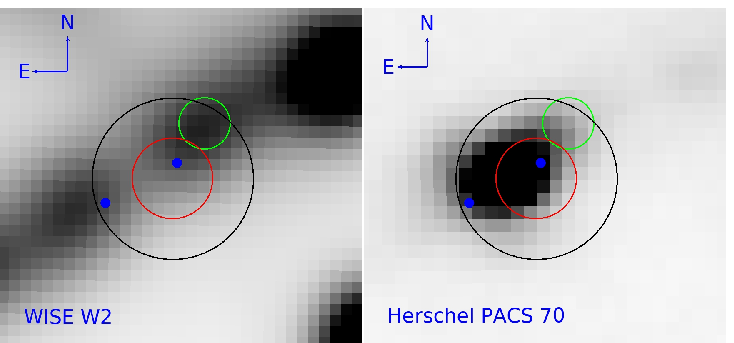}}
    \caption{(top, left) {\it WISE} $W1$ (blue circles) and $W2$ (red circles) light curves compared with the 850 $\mu$m flux (black circles) obtained by the JCMT survey for source 0 in IC348. We note that the apparent multiple data points over single dates in the $W1$ and $W2$ light curves are just an effect of the cadence of {\it WISE} observations (several visits over $\sim$ 1 day). For clarity, we only show the portion of the data that has nearly contemporaneous observations. For JCMT fluxes we also show the mean peak brightness (horizontal black dashed line), the fiducial standard deviation (SD$_{{\rm fid}}$, see main text, blue dot-dashed line) and the measured standard deviation (red dot-dashed line). (top, right) 12$\times$12 arcsec area around the location of the JCMT source (marked by the red circle). In the plot, we also show 5 arcsec (red) and 10 arcsec (black) circles that represent the error in the location of the JCMT source and the search radius for {\it WISE} counterparts, respectively. Known YSOs are marked by the blue filled circles. The location of the multi-epoch {\it WISE} and {\it NEOWISE} data found within the 10 arcsec search radius are marked by the small black points. Green diamonds show the location of the detections that were actually used to build $W1$ and $W2$ light curves presented in the top left panel (see main text). The green 6.4 arcsec diameter circle around the median $\alpha$ and $\delta$ of {\it WISE} detections is shown to represent the angular resolution of {\it WISE}.  (bottom) {\it WISE} $W2$ (left) and {\it Herschel} PACS 70 $\mu$m \citep{2010Poglitsch} 2$\times2^\prime$~ images around the location of the JCMT source. The red, blue and green circles are the same that are presented in the top-right plot.}
    \label{fig:ex1}
\end{figure*}

In order to study the effects of variable accretion during the embedded phase of stellar evolution we began a long-term monitoring program of the sub-millimetre flux of YSOs in eight known star forming regions, using SCUBA-2 at the James Clerk Maxwell Telescope \citep[the JCMT Transient Survey;][]{2017Herczeg}. The programme has detected robust variability at sub-millimetre wavelengths. \citet{2018Johnstone} found several secular and stochastic variable protostars using data from the first 18 months of the JCMT survey. As well, \citet{2017Mairs} compared the first epochs of the Transient Survey against observations of the same regions taken $\sim$5 years earlier as part of the Gould Belt Survey \citep[GBS;][]{2007Ward} and found a similar number of variables. In most cases the same sources were found to vary both on long timescales, $\sim$5 years, and short timescales, 18 months. These include the class I protostar EC 53 in the Serpens Main region \citep{2017Yoo}, a known near-IR periodic variable where changes in the accretion rate are suggested to be the main driver of the variability of the source \citep{2012Hodapp}.

Monitoring from the JCMT GBS and Transient surveys, along with contemporaneous observations at near- and mid-IR wavelengths \citep[from multi-epoch UKIDSS GPS and/or {\it WISE} surveys][]{2017Lucas, 2010Wright}, provides us with the opportunity to study changes in the spectral energy distribution (SED) over a wide wavelength range. Near- and mid-IR photometric surveys have found that variability is common amongst YSOs, with objects at earlier evolutionary stages showing the largest amplitudes \citep[see e.g.][]{c2011Morales,2014Rebull,2017Contreras_a,2017Lucas,2018Wolk}. At shorter wavelengths, physical mechanisms such as cold spots in the stellar photosphere, hot spots due to accretion, or variable extinction due to a warped inner disc \citep[][]{2013Bouvier} can all explain the observed variability in most YSOs \citep[e.g.][]{2014Cody,sergison2020}. Given that these mechanisms occur at the star-inner disc interface, the timescales of the variability are in the order of days, and do not lead to longer-term (months to years) variability. The latter is likely caused by variable accretion (as we have discussed previously) or changes in the extinction along the line of sight due to structures at larger radii in the disc \citep[see e.g. the long-term obscuration event in AA Tau;][]{2013Bouvier}.

Detecting variability at far-IR and sub-mm wavelengths has the advantage that extinction changes or asymmetric hot or cold-spots cannot lead to variability at these wavelengths. The most likely cause is a large change in the accretion rate of the system (e.g \citealt{2013Johnstone}; \citealt{2019Macfarlane_a}). Comparisons between changes at sub-mm (or far-IR) and near-IR wavelengths, however,  have usually been limited to individual objects with outburst and pre- or post-outburst SEDs \citep[e.g.][]{2007Kospal,Juhasz2012,2013kospal,2015Safron}.

To aid our understanding of the structure of envelopes in young stellar objects, we study in this paper, for the first time, the continuum variability of a sample of deeply embedded sources by comparing the ongoing observations of the JCMT Transient Survey with the all-sky mid-IR photometry from {\it WISE}/ {\it NEOWISE} \citep{2010Wright, 2011Mainzer, 2013Cutri}. Section \ref{sec:surveys} describes the data from the surveys used on this work. In Section \ref{sec:method} we describe how we crossmatched the data from both surveys and explain the different steps that were taken to arrive to our final sample of 59 YSOs. Section \ref{sec:statvar}
 defines the statistical measurements of stochastic and secular variability used to analyse our sample. In Section \ref{sec:wisevar} we focus on finding the correlation between the secular changes at $4.6$ and $850 \mu$m. Finally, in Section \ref{sec:wjvar} we study the mid-IR to sub-mm continuum variability of embedded YSOs using 14 objects which follow a similar correlation between variability at 4.6 $\mu$m and 850$\mu$m. In this Section we also discuss how different properties of the YSOs can impact this relation.

\section{Surveys}\label{sec:surveys}

\setlength{\tabcolsep}{2.5pt}
\begin{table*}
	\centering
	\caption{The 59 sources where JCMT and {\it WISE} fluxes likely arise from the same source.}
	\label{tab:allvar}
\resizebox{\textwidth}{!}{
   	\begin{tabular}{lccccccccccccccccc} 
		\hline
		Region & Source  & $\alpha$ & $\delta$ & JCMT designation & Name & YSO class & YSO reference & $\overline{W2}$ & $\Delta W2$ ($90\%$) & $\Delta W2 $ & SD/SD$_{\rm fid}$ & $S/\Delta S$  & N$_{\rm W2}$ & $\overline{Flux}$  &  SD/SD$_{fid}$  & $S/\Delta S$ & N$_{\rm JCMT}$ \\
		             &                           &                &               &                              &           &                   &                          &    (mag)                &             (mag)                 &         (mag)       &     ($W2$)                     &     ($W2$)                                                          &                   &          (Jy beam$^{-1}$)            &     (JCMT)             &       (JCMT)         &                      \\                          
		\hline
IC348 &        1 & 03:43:57.0 & $$+$$32:03:04.9 & JCMTPP\_J034357.0$+$320305 & [CAZ2013] IC348MMS1 & I & \citet{2013Dunham} & 12.29 &  0.7 &  0.9 &  1.60 &  11.81 & 13 &  1.20 &  1.54 &   6.31 & 27 \\
IC348 &        2 & 03:43:50.9 & $$+$$32:03:22.8 & JCMTPP\_J034350.9$+$320323 & SSTc2d J034351.0$+$320325 & I & \citet{2013Dunham} & 11.59 &  0.4 &  0.5 &  1.80 &  10.90 & 15 &  0.34 &  0.99 &   0.54 & 28 \\
IC348 &       14 & 03:44:12.8 & $$+$$32:01:35.0 & JCMTPP\_J034412.8$+$320135 & SSTc2d J034413.0$+$320135 & F & \citet{2012Kryukova} &  6.90 &  0.9 &  1.1 &  3.00 & $-$14.94 & 16 &  0.16 &  1.15 &   0.63 & 28 \\
NGC1333 &        0 & 03:29:10.4 & $$+$$31:13:30.9 & JCMTPP\_J032910.4$+$311331 & [JCC87] IRAS4A & 0 & \citet{2013Dunham} & 14.27 &  0.8 &  0.8 &  0.10 &  $-$1.81 &  4 &  9.15 &  1.73 &  $-$7.20 & 31 \\
NGC1333 &        3 & 03:28:55.6 & $$+$$31:14:34.0 & JCMTPP\_J032855.6$+$311434 & SSTc2d J032855.6$+$311437 & I & \citet{2013Dunham} &  9.98 &  0.7 &  0.7 &  0.80 &   0.86 &  7 &  2.45 &  1.61 &   0.59 & 30 \\
NGC1333 &        8 & 03:29:03.8 & $$+$$31:14:49.0 & JCMTPP\_J032903.8$+$311449 & SSTc2d J032904.1$+$311447 & I & \citet{2013Dunham} & 11.95 &  0.5 &  0.5 &  1.30 &   8.59 &  8 &  0.47 &  1.70 &   6.75 & 25 \\
NGC1333 &       17 & 03:29:11.1 & $$+$$31:18:27.9 & JCMTPP\_J032911.1$+$311828 & ASR 32 & I & \citet{2013Dunham} & 10.39 &  0.2 &  0.2 &  1.20 &  $-$7.26 & 11 &  1.11 &  0.94 &  $-$2.91 & 25 \\
NGC1333 &       18 & 03:29:01.5 & $$+$$31:20:28.0 & JCMTPP\_J032901.5$+$312028 & SSTc2d J032901.6$+$312021 & I & \citet{2013Dunham} &  5.81 &  0.6 &  0.7 &  1.10 &  $-$3.16 & 12 &  1.06 &  1.05 &  $-$1.25 & 30 \\
NGC1333 &       23 & 03:29:13.4 & $$+$$31:18:09.9 & JCMTPP\_J032913.4$+$311810 & SSTc2d J032913.0$+$311814 & I & \citet{2013Dunham} &  7.71 &  0.1 &  0.2 &  0.70 &  $-$0.48 & 12 &  0.32 &  1.19 &  $-$0.48 & 30 \\
NGC1333 &       24 & 03:29:07.8 & $$+$$31:21:55.0 & JCMTPP\_J032907.8$+$312155 & [LAL96] 213 & 0 & \citet{2013Dunham} &  6.75 &  0.3 &  2.7 &  0.10 &  $-$3.59 &  4 &  0.33 &  2.06 &  $-$3.55 &  6 \\
NGC1333 &       29 & 03:28:56.3 & $$+$$31:19:13.0 & JCMTPP\_J032856.3$+$311913 & SSTc2d J032856.1$+$311908 & II & \citet{2015Young} &  9.67 &  0.1 &  0.2 &  0.60 &   0.34 & 12 &  0.16 &  0.92 &   0.56 & 30 \\
NGC1333 &       35 & 03:28:36.9 & $$+$$31:13:27.9 & JCMTPP\_J032836.9$+$311328 & SSTc2d J032837.1$+$311331 & I & \citet{2013Dunham} &  8.81 &  0.7 &  0.8 &  1.70 & $-$11.37 & 15 &  0.46 &  1.18 &  $-$1.44 & 31 \\
NGC1333 &       37 & 03:29:17.2 & $$+$$31:27:45.9 & JCMTPP\_J032917.2$+$312746 & SSTc2d J032917.2$+$312746 & I & \citet{2013Dunham} & 12.85 &  0.3 &  0.9 &  0.70 &  $-$7.65 & 18 &  0.31 &  1.16 &  $-$0.02 & 31 \\
NGC1333 &       43 & 03:28:34.6 & $$+$$31:07:03.9 & JCMTPP\_J032834.6$+$310704 & SSTc2d J0328345$+$310705 & I(?) & \citet{2013Dunham} & 10.61 &  0.6 &  0.9 &  5.20 &  17.65 & 15 &  0.23 &  1.41 &   1.53 & 31 \\
NGC2024 &        6 & 05:41:36.0 & $$-$$01:56:24.0 & JCMTPP\_J054136.0$-$015624 & 2MASS J05413581$-$0156222 & II & \citet{2012Megeath} &  8.10 &  0.3 &  0.3 &  0.10 &  $-$0.10 &  2 &  0.16 &  0.78 &  $-$1.38 & 29 \\
NGC2024 &        7 & 05:41:41.2 & $$-$$01:58:00.0 & JCMTPP\_J054141.2$-$015800 & 2MASS J05414164$-$0157545 & I & \citet{2012Kryukova} &  7.22 &  0.2 &  0.2 &  0.90 &  $-$0.13 & 12 &  0.16 &  1.22 &   1.02 & 29 \\
NGC2024 &       22 & 05:42:02.6 & $$-$$02:07:39.0 & JCMTPP\_J054202.6$-$020739 & HOP S303 & I & \citet{2012Kryukova} & 11.53 &  0.2 &  0.2 &  1.00 &  $-$1.99 & 15 &  0.81 &  1.40 &  $-$1.38 & 30 \\
NGC2024 &       59 & 05:41:36.0 & $$-$$01:37:42.0 & JCMTPP\_J054136.0$-$013742 & WISE J054135.67$-$013748.1 & I(?) & \citet{2012Cutri} & 13.54 &  0.3 &  0.3 &  0.30 &   0.23 & 10 &  0.24 &  1.10 &  $-$0.72 & 29 \\
NGC2024 &       62 & 05:41:27.4 & $$-$$01:47:54.0 & JCMTPP\_J054127.4$-$014754 & -- & I(?) & -- & 12.13 &  0.5 &  0.6 &  0.30 &  $-$0.20 &  7 &  0.18 &  0.87 &  $-$0.54 & 29 \\
NGC2068 &        0 & 05:46:08.4 & $$-$$00:10:41.0 & JCMTPP\_J054608.4$-$001041 & HOPS 317 & 0 & \citet{2016Furlan} &  9.93 &  0.3 &  0.4 &  1.50 &   7.44 & 11 &  2.58 &  0.97 &   3.08 & 29 \\
NGC2068 &        1 & 05:46:07.2 & $$-$$00:13:32.0 & JCMTPP\_J054607.2$-$001332 & HOPS 358 & 0 & \citet{2016Furlan} &  9.60 &  0.9 &  0.9 &  6.20 & $-$31.52 &  9 &  1.31 &  3.33 & $-$12.16 & 24 \\
NGC2068 &        2 & 05:46:08.2 & $$-$$00:09:59.0 & JCMTPP\_J054608.2$-$000959 & HOPS 386 & I & \citet{2016Furlan} &  6.05 &  1.0 &  1.1 &  2.10 &  $-$2.20 & 13 &  0.62 &  0.96 &   1.25 & 30 \\
NGC2068 &        4 & 05:46:03.6 & $$-$$00:14:47.0 & JCMTPP\_J054603.6$-$001447 & HOP S315 & I & \citet{2016Furlan} &  7.40 &  0.4 &  0.5 &  1.90 &  12.24 & 15 &  0.52 &  0.78 &   1.50 & 30 \\
NGC2068 &        5 & 05:46:07.6 & $$-$$00:11:50.0 & JCMTPP\_J054607.6$-$001150 & [FM2008] 458 & II & \citet{2008Flaherty} &  7.86 &  0.0 &  0.0 &  0.20 &   0.06 & 15 &  0.53 &  1.02 &   3.00 & 30 \\
NGC2068 &       12 & 05:46:31.0 & $$-$$00:02:32.0 & JCMTPP\_J054631.0$-$000232 & HOPS 373 & 0 & \citet{2016Furlan} & 10.91 &  0.3 &  0.8 &  1.10 & $-$10.05 & 16 &  1.23 &  1.51 &  $-$4.88 & 30 \\
NGC2068 &       14 & 05:46:47.4 & $$+$$00:00:28.0 & JCMTPP\_J054647.4$+$000028 & HOPS 323 & I & \citet{2016Furlan} &  7.92 &  0.5 &  0.5 &  1.70 &  $-$6.67 & 10 &  1.00 &  1.29 &  $-$4.22 & 24 \\
NGC2068 &       15 & 05:46:37.8 & $$+$$00:00:37.0 & JCMTPP\_J054637.8$+$000037 & HOPS 324 & I & \citet{2016Furlan} &  9.26 &  0.8 &  1.1 &  7.80 &   6.54 & 12 &  0.60 &  1.30 &   3.82 & 30 \\
NGC2068 &       23 & 05:46:33.4 & $$-$$00:00:05.0 & JCMTPP\_J054633.4$-$000005 & HOPS 321 & I & \citet{2016Furlan} &  9.36 &  0.2 &  0.3 &  2.70 &  14.11 & 13 &  0.30 &  1.07 &  $-$0.15 & 30 \\
OMC 2/3 &        6 & 05:35:23.4 & $$-$$05:12:02.0 & JCMTPP\_J053523.4$-$051202 & HOPS 60 & 0 & \citet{2016Furlan} &  7.94 &  0.4 &  0.5 &  1.40 &  $-$4.78 & 12 &  1.38 &  1.11 &  $-$0.23 & 27 \\
OMC 2/3 &        9 & 05:35:23.4 & $$-$$05:07:05.0 & JCMTPP\_J053523.4$-$050705 & 2MASS J05352332$-$0507096 & F & \citet{2012Megeath} &  9.03 &  0.2 &  0.3 &  1.20 &  $-$4.22 & 12 &  1.06 &  1.14 &  $-$0.46 & 27 \\
OMC 2/3 &       17 & 05:35:14.9 & $$-$$05:16:08.0 & JCMTPP\_J053514.9$-$051608 & -- & I(?) & -- &  9.79 &  0.3 &  0.4 &  0.60 &   0.75 &  8 &  0.38 &  0.93 &  $-$1.51 & 27 \\
OMC 2/3 &       22 & 05:35:14.9 & $$-$$05:16:38.0 & JCMTPP\_J053514.9$-$051638 & [H97b] 20475 & II & \citet{2013Broos} & 10.72 &  0.1 &  0.2 &  0.20 &  $-$0.74 &  8 &  0.27 &  1.20 &  $-$1.88 & 27 \\
OMC 2/3 &       30 & 05:35:18.1 & $$-$$05:13:35.0 & JCMTPP\_J053518.1$-$051335 & [CHS2001] 9147 & F & \citet{2012Megeath} &  6.80 &  0.4 &  0.5 &  1.40 &   2.79 & 12 &  0.18 &  1.14 &  $-$0.45 & 27 \\
OMC 2/3 &       47 & 05:35:15.9 & $$-$$04:59:56.0 & JCMTPP\_J053515.9$-$045956 & [CHS2001] 8787 & I & \citet{2012Kryukova} &  7.49 &  0.3 &  0.5 &  1.80 &   8.47 & 15 &  1.15 &  2.10 &   5.11 & 16 \\
OMC 2/3 &       52 & 05:35:29.8 & $$-$$04:59:44.0 & JCMTPP\_J053529.8$-$045944 & HOPS 383 & 0 & \citet{2016Furlan} & 11.90 &  1.0 &  1.5 &  3.60 & $-$10.83 &  5 &  0.56 &  1.62 &  $-$5.77 & 19 \\
OMC 2/3 &       88 & 05:35:14.5 & $$-$$05:18:41.0 & JCMTPP\_J053514.5$-$051841 & 2MASSJ 05351467$-$0518433 & II & \citet{2013Broos} &  7.87 &  0.2 &  0.2 &  1.00 &  $-$2.26 &  6 &  3.59 &  1.07 &  $-$2.01 & 27 \\
OMC 2/3 &       92 & 05:35:20.6 & $$-$$05:19:17.0 & JCMTPP\_J053520.6$-$051917 & [AD95]1362 & I & \citet{2012Megeath} &  7.20 &  0.4 &  0.7 &  0.10 &  $-$2.89 &  4 &  0.96 &  1.28 &  $-$1.10 & 27 \\
OMC 2/3 &      143 & 05:34:29.4 & $$-$$04:55:28.8 & JCMTPP\_J053429.4$-$045529 & HOPS 99 & 0 & \citet{2016Furlan} & 12.22 &  0.2 &  0.3 &  0.40 &   0.98 & 12 &  0.21 &  0.96 &  $-$1.40 & 27 \\
Ophiuchus Core &       36 & 16:27:05.4 & $$-$$24:36:28.0 & JCMTPP\_J162705.4$-$243628 & [EDJ2009] 862 & I & \citet{2012Kryukova} & 10.15 &  0.2 &  0.3 &  0.60 &  $-$4.63 & 11 &  0.18 &  1.32 &  $-$1.54 & 23 \\
Ophiuchus Core &       42 & 16:26:40.8 & $$-$$24:27:15.9 & JCMTPP\_J162640.8$-$242716 & [EDJ2009] 831 & F & \citet{2012Kryukova} &  8.65 &  0.9 &  1.0 &  3.30 & $-$11.36 & 11 &  0.24 &  1.11 &  $-$0.93 & 23 \\
Ophiuchus Core &       45 & 16:26:44.3 & $$-$$24:34:48.9 & JCMTPP\_J162644.3$-$243449 & [EDJ2009] 836 & I & \citet{2012Kryukova} &  6.52 &  0.9 &  1.1 &  2.30 &  12.27 & 11 &  0.20 &  1.49 &  $-$1.24 & 23 \\
Ophiuchus Core &       48 & 16:28:21.6 & $$-$$24:36:23.8 & JCMTPP\_J162821.6$-$243624 & [EDJ2009] 954 & I & \citet{2012Kryukova} & 12.19 &  0.8 &  0.9 &  4.20 & $-$14.64 & 12 &  0.22 &  1.33 &  $-$0.39 & 23 \\
Ophiuchus Core &       56 & 16:28:16.7 & $$-$$24:36:56.9 & JCMTPP\_J162816.7$-$243657 & WSB 60 & II & \citet{2013Dunham} &  7.90 &  0.2 &  0.3 &  1.20 &   1.34 & 11 &  0.17 &  1.90 &  $-$0.19 & 23 \\
Serpens Main &        0 & 18:29:49.8 & $$+$$01:15:20.0 & JCMTPP\_J182949.8$+$011520 & Serpens SMM1 & 0 & \citet{2013Dunham} &  9.44 &  0.9 &  1.0 &  4.50 &  13.48 &  8 &  7.00 &  2.28 &   8.63 & 28 \\
Serpens Main &        1 & 18:29:48.2 & $$+$$01:16:44.0 & JCMTPP\_J182948.2$+$011644 & SSTc2d J182948.1$+$011644 & I & \citet{2012Kryukova} & 10.94 &  0.2 &  0.3 &  1.20 &  $-$3.44 & 14 &  2.11 &  0.86 &  $-$2.54 & 41 \\
Serpens Main &        2 & 18:29:51.2 & $$+$$01:16:38.0 & JCMTPP\_J182951.2$+$011638 & EC 53 & I & \citet{2013Dunham} &  7.94 &  1.9 &  1.9 & 10.10 &  25.27 &  7 &  1.19 &  4.50 &  17.33 & 32 \\
Serpens Main &        3 & 18:29:52.0 & $$+$$01:15:50.0 & JCMTPP\_J182952.0$+$011550 & Serpens SMM10IR & I & \citet{2012Kryukova} &  8.99 &  0.9 &  1.0 &  5.10 &  29.23 &  9 &  0.84 &  1.58 &   6.21 & 24 \\
Serpens South &       11 & 18:29:59.6 & $$-$$02:01:00.0 & JCMTPP\_J182959.6$-$020100 & MHO 3271 & I & \citet{2015Dunham} & 11.94 &  0.7 &  0.7 &  0.10 & $-$12.65 &  4 &  0.29 &  1.47 &   3.23 & 26 \\
Serpens South &       13 & 18:30:01.0 & $$-$$02:06:12.0 & JCMTPP\_J183001.0$-$020612 & 2MASS J18300101$-$0206082 & I & \citet{2015Dunham} &  9.52 &  0.2 &  0.2 &  3.50 &   1.46 &  8 &  0.20 &  0.82 &   0.27 & 25 \\
Serpens South &       15 & 18:30:16.0 & $$-$$02:07:21.0 & JCMTPP\_J183016.0$-$020721 & MHO 3274 & I & \citet{2015Dunham} & 11.69 &  0.2 &  0.3 &  1.40 &  $-$1.08 &  8 &  0.20 &  1.02 &  $-$0.35 & 25 \\
Serpens South &       20 & 18:29:47.0 & $$-$$01:55:54.0 & JCMTPP\_J182947.0$-$015554 & IRAS1 8271$-$0157 & F & \citet{2013Dunham} &  8.54 &  0.4 &  0.4 &  2.50 &  $-$5.32 &  8 &  0.16 &  1.22 &  $-$0.96 & 25 \\
Serpens South &       36 & 18:31:10.2 & $$-$$02:06:44.9 & JCMTPP\_J183110.2$-$020645 & SSTU J183110.35$-$020637.0 & I & \citet{2013Mallick} &  9.97 &  0.2 &  0.3 &  1.50 &  $-$1.83 &  6 &  0.64 &  1.33 &  $-$0.37 & 25 \\
Serpens South &       47 & 18:29:41.8 & $$-$$01:50:21.0 & JCMTPP\_J182941.8$-$015021 & SSTgbs J1829419$-$015011 & I & \citet{2015Dunham} & 11.71 &  0.4 &  0.4 &  2.00 &  $-$1.56 &  9 &  0.50 &  1.14 &   0.29 & 26 \\
Serpens South &       54 & 18:30:25.8 & $$-$$02:10:45.0 & JCMTPP\_J183025.8$-$021045 & 2MASS J18302593$-$0210420 & I & \citet{2013Dunham} &  8.27 &  0.1 &  0.1 &  0.10 &   0.90 &  4 &  0.45 &  1.18 &   0.38 & 25 \\
Serpens South &       55 & 18:30:28.8 & $$-$$01:56:06.0 & JCMTPP\_J183028.8$-$015606 & MHO 3279 & I & \citet{2013Povich} & 12.15 &  0.3 &  0.4 &  1.80 &  $-$0.67 &  9 &  0.26 &  1.13 &   0.02 & 26 \\
Serpens South &       58 & 18:30:49.2 & $$-$$01:56:06.0 & JCMTPP\_J183049.2$-$015606 & MHO 3281 & I & \citet{2015Dunham} &  9.77 &  0.3 &  0.5 &  6.60 & $-$12.16 &  6 &  0.23 &  1.02 &  $-$0.82 & 26 \\
Serpens South &       70 & 18:29:12.8 & $$-$$02:03:54.0 & JCMTPP\_J182912.8$-$020354 & MSX6C G028.5532$+$03.9958 & I & \citet{2015Dunham} &  6.50 &  0.8 &  0.9 &  3.50 &  $-$6.99 &  7 &  0.16 &  1.07 &  $-$0.13 & 25 \\
Serpens South &       73 & 18:29:43.2 & $$-$$01:56:51.0 & JCMTPP\_J182943.2$-$015651 & [MAM2011] SerpS-MM4 & I & \citet{2013Dunham} &  8.12 &  0.3 &  0.4 &  1.50 &   1.30 &  8 &  0.16 &  0.99 &   0.79 & 26 \\
Serpens South &       74 & 18:29:43.8 & $$-$$02:12:57.0 & JCMTPP\_J182943.8$-$021257 & [ZFW2015] 12 & I & \citet{2013Dunham} & 11.87 &  0.2 &  0.2 &  1.40 &  $-$2.74 &  6 &  0.16 &  1.18 &   0.48 & 25 \\
		\hline
 	\end{tabular}}
\end{table*} 

\setlength{\tabcolsep}{6pt}

\subsection{JCMT Transient Survey}\label{ssec:jcmt_desc}

The JCMT Transient Survey uses the SCUBA-2 instrument \citep{2013Holland} on JCMT to monitor sub-mm continuum emission from eight nearby star-forming regions.  The eight regions, Ophiuchus Core, NGC 1333, IC 348, Serpens Main, Serpens South, OMC 2/3, NGC 2024, NGC 2068, were selected for the high density of deeply embedded protostars \citep[][]{2017Herczeg}.  Each region is observed in a PONG mode that produces an image with smooth sensitivity across a field with $30^\prime$ diameter, with an integration time set to reach $\sim 12$ mJy at 850 $\mu$m.  Some epochs have low enough precipitable water vapour (PWV) to also image the region at 450 $\mu$m. 
The data are reduced using customized routines, including spatial masks and offsets, with the map-making 
software, {\sc{makemap}} \citep[see ][ for details]{2013Chapin} in the {\sc{starlink}} package \citep{2013Jenness,2014Currie}.  The 850 $\mu$m fluxes are measured from the peak brightness of the object, and are then calibrated using bright sources that are measured to be non-varying at 850 $\mu$m.

A full description of our reductions and calibrations are described by \citet{2017Mairsb}.  The uncertainty most relevant to this paper is the flux calibration uncertainty in any single epoch of $\sim 0.025$ F$_{850}$+12 [mJy], determined by a combination of the noise level and the stability of calibrator sources.  The  uncertainty in the absolute spatial position is $\sim 3^{\prime\prime}$, which affects our ability to match sources across different surveys.  The beam size of SCUBA-2 is 14.6$^{\prime\prime}$ at 850~$\mu$m.

\subsection{{\it WISE}/{\it NEOWISE}}\label{sec:wise_exp}

The Wide Infrared Survey Explorer \citep[{\it WISE}, ][]{2010Wright} is a 40 cm telescope in a low-earth orbit that surveyed the entire sky in 2010 using four infrared bands centred at 3.4, 4.6, 12 and 22 $\mu$m (denoted $W1, W2, W3$ and $W4$ respectively) and with an angular resolution of 6.1$^{\prime\prime}$, 6.4$^{\prime\prime}$, 6.5$^{\prime\prime}$ and 12.0$^{\prime\prime}$, respectively.  The orbit of {\it WISE} allowed it to cover every part of the sky at least eight times \citep{2011Mainzer}, with each patch of sky observed many times over a period of $\sim$ a day. The survey ran between January and September 2010, when the telescope's cryogen tanks were depleted. After this time, the telescope continued to operate for four months using the $W1$ and $W2$ bands, and with the same original survey strategy \citep[known as the {\it NEOWISE} Post-Cryogenic Mission,][]{2011Mainzer}. With the primary aim of studying near-Earth objects, the {\it NEOWISE} mission was reactivated in 2013 \citep{2014Mainzer}  and has continued to operate with the latest data release containing observations through mid-December 2018. 

In this work we used the $W1$ and $W2$ observations from the WISE All-Sky single exposure database \citep{2012Cutri}, which contains observations taken between January and August 2010. In addition we used the NEOWISE single exposure database (2019 data release) that contains $W1$ and $W2$ observations from December 2013 until December 2018 \citep[][]{2015Cutri}. Single {\it WISE} exposures saturate at $W1\sim7.8$ and $W2\sim6.8$~mag \citep{2012Cutri}, while the NEOWISE single-exposure detections are complete up to $W1=15$ and $W2=13$~mag \citep{2015Cutri}.

\begin{figure}
	\resizebox{1\columnwidth}{!}{\includegraphics{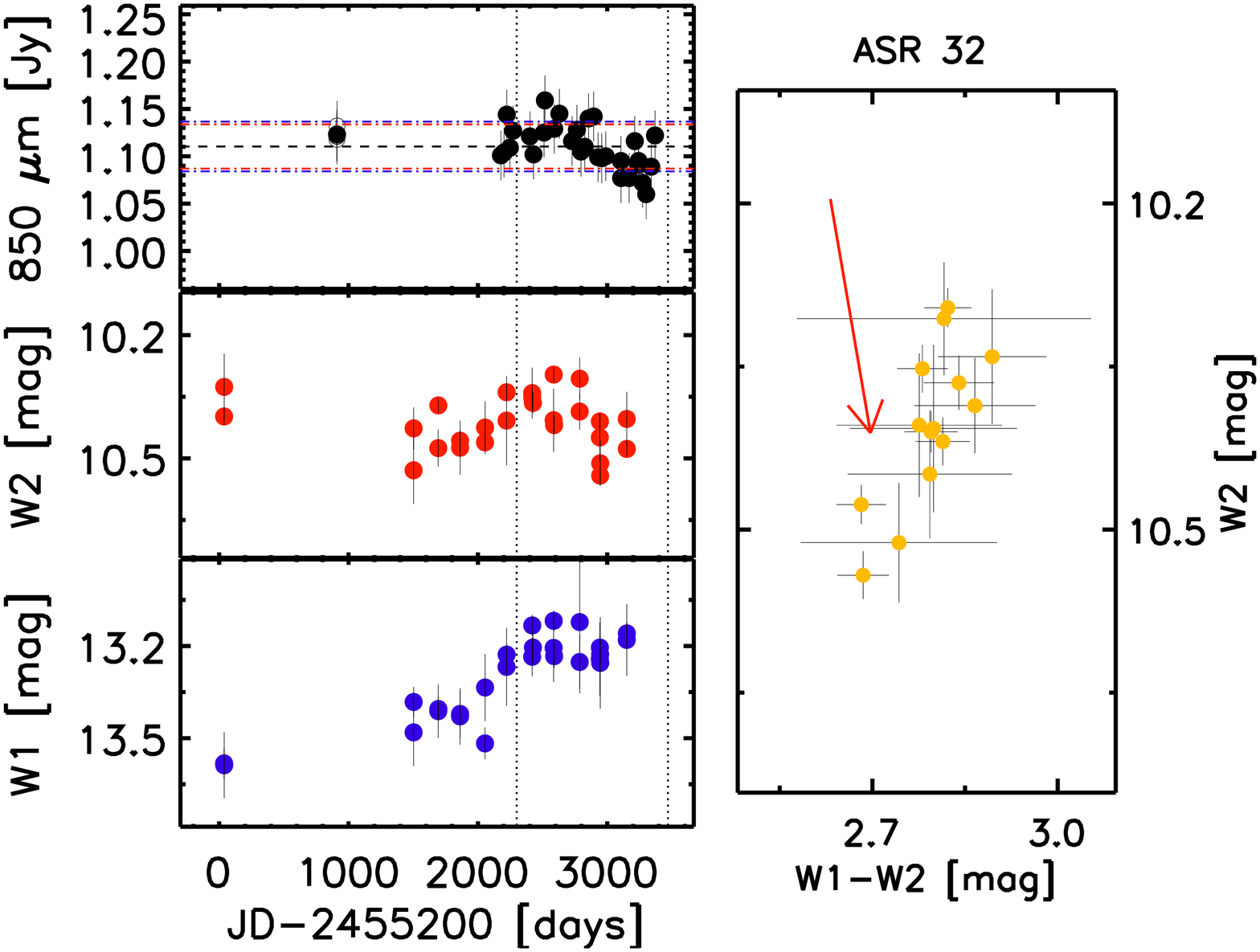}}
	 \caption{(left) {\it WISE} $W1$ (blue circles), $W2$ (red circles) magnitudes, and 850 $\mu$m flux from the JCMT for source 17 in NGC1333. Horizontal lines are the same as in Fig 1. In the plots the black vertical dashed lines mark the time limits that were used to search for correlated variability between the {\it WISE} and JCMT data. (right) $W2$ vs $W1-W2$ for the {\it WISE} data taken within the defined time limits. In the plot, the red arrow indicates the reddening vector with $A_{\rm K}=0.5$~mag, adapted from \citet{2005Indebetouw}.}
    \label{fig:n1333_17}
\end{figure}

\section{Method}\label{sec:method}

Several steps were taken to produce a catalogue of YSOs with JCMT and {\it WISE} detections to understand the relationship between the mid-IR to sub-mm continuum variability. These steps are described below.

\subsection{The initial JCMT sample}\label{ssec:initjcmt}

The JCMT Transient survey has 1665 sources at 850 $\mathrm{\mu}$m, including protostars, prestellar cores, and starless clumps. This work, however, will only focus on those objects which are associated with protostars and that are bright enough to detect an accretion-driven outburst in the JCMT data.

Using a simple model of a central protostar embedded in a spherically symmetric envelope \citet{2013Johnstone} find that enhancements in the accretion luminosity lead to a flux increase across the whole spectrum of an embedded YSO. SED models that include the contribution of an accretion disc and inner cavities due to outflows also find similar results \citep[see e.g.][]{2013Scholz, 2019Macfarlane_a, 2020Baek}. The emission at long wavelengths responds to the changing dust temperature in the outer envelope, whereas the mid-IR wavelengths should respond to luminosity changes from the inner disc and protostar.   We thus anticipate a relation between the mid-IR and sub-mm with $\log$ flux$_{\rm IR}/\log$ flux$_{850} \propto 4-6$ (see also Section \ref{sec:wjvar}).

In the search for accretion-driven YSO outbursts in the mid-IR \citet{2013Scholz} adopted an amplitude cut-off of $\Delta 3.4 (4.6) \mathrm{\mu m} > 1$~mag. This avoids selecting YSOs with variability being driven by other common mechanisms affecting the stellar photosphere or the star-disc interface such as e.g. hot spots or a warped inner disc. If we apply a similar cutoff, and given the expected relationship between the IR and sub-mm fluxes estimated above, an amplitude of 1 magnitude at 4.6 $\mathrm{\mu m}$ (or a change by a factor of 2.5 in flux) corresponds to a change by a factor of 1.25 at 850 $\mu$m . Then, an accretion burst of 1 mag in {\it WISE} would cause a 0.1 Jy beam$^{-1}$ source to brighten to 0.125 Jy beam$^{-1}$, a difference that is less than twice the noise of 0.014 Jy beam$^{-1}$ in any single epoch \citep{2017Mairsb}.

Therefore, we will analyse only JCMT sources with S/N$>10$, corresponding to a mean brightness greater than or equal to 0.15 Jy beam$^{-1}$). This selection reduces the sample to 307 JCMT sources.

\subsection{Selection from {\it WISE}}

We searched for {\it WISE} and  {\it NEOWISE} counterparts for all of the 307 bright JCMT sources using a 10 arcsec radius. For both mid-IR surveys we queried the single exposure source databases at the NASA/IPAC infrared science archive. To make sure that the $W1$ and $W2$ magnitudes arise from the same source, we calculated the median and standard deviation of the right ascension ($\alpha$) and declination ($\delta$) of all of the detections within the search radius. Then we only considered detections that were within 1$\sigma$ of the median $\alpha$ and $\delta$. Finally, since several {\it WISE} exposures may be obtained within a few hours of each other, we produced a catalogue with the mean MJD, mean magnitude, and error by combining the mean of the error in single exposures and the standard deviation of the magnitudes taken during the same day. 

In addition we made the following cuts before including a source in the final statistics sample.

\begin{itemize}  
\item To avoid including faint sources or objects that are saturated in both bands, only sources with  $15>W1>7.8$~mag or $13>W2>6.8$~mag are included in this analysis (see Section \ref{sec:wise_exp}).
\item  Objects need to have at least 5 detections (within the period of time where both JCMT and {\it WISE} are contemporaneous) in both $W1$ and $W2$, 5 detections in $W2$ when not detected in $W1$ or 5 detections in $W1$ when not detected in $W2$. The number of data points is selected to keep a reasonable sample while also demanding a meaningful number of data points.

\end{itemize}

A total of 126 sources fulfil these criteria.  In some cases, objects that fulfil the magnitude criteria in $W1$ fall in the saturated regime in $W2$. In these cases {\it WISE} $W2$ fluxes are corrected for saturation following the guidance from the {\it WISE} supplementary material \citep{2012Cutri}. 

Of the 181 JCMT sources that fail the above criteria, 81  are not detected in {\it WISE}, 20 are too faint (or not detected) in $W1$ or $W2$, and 56 sources have fewer than 5 detection in both {\it WISE} bands. In addition, 24 JCMT sources are saturated in the {\it WISE} photometry. Of the 157 objects with no detections/too faint/low number of detections,  138 are not likely to be associated with known protostars, i.e. they are not associated with known YSOs  \citep[from the near- and mid-IR photometric catalogues of][]{2013Stutz, 2012Megeath, 2015Dunham} within a 10 arcsec radius from the JCMT source.   These sub-mm peaks that lack mid-IR counterparts may be pre-stellar cores of possibly very young Class 0 sources.

Of the 24 objects that are saturated in {\it WISE},  21 are associated with known protostars. We present a search for potential variability for these sources in Appendix \ref{app:sat}. However, since these objects suffer from saturation in the {\it WISE} bands we will not discuss them in the main section of the paper.

\subsection{Visual Inspection}\label{sec:visual}

To understand whether a correlation is real or not, we needed to inspect the {\it WISE} light curves and images to determine the reliability of the photometric data by confirming that the flux measured at both wavelengths corresponds to the same source. Since we used a large radius, and given the large beam size of the JCMT data (FWHM of 14.6 arcsec), there could be cases where several YSOs are found within the search radius, which leads to the possibility of the mid-IR and 850$\mu$m emission arising from different sources.

We visually inspected the {\it WISE} images and photometry for the 126  matched sources. In the images we also compared the coordinates of the JCMT and WISE sources with those of YSOs found in the 10 arcsec radius \citep[from the][catalogues]{2013Stutz, 2012Megeath, 2015Dunham}. In many cases we also used images from {\it Herschel} \citep{2010Pilbratt} to determine the likelihood of the WISE and JCMT detections corresponding to the same source. 

For example, for source 0 in the IC 348 region \citep[SSTc2d J0343565$+$320052][]{2013Dunham},  the location of the WISE source does not correspond to either of the two known YSOs in the area. Further inspection of images shows that the WISE source is very likely not responsible for the emission at longer wavelengths, as it becomes fainter at 70 $\mu$m (Fig. \ref{fig:ex1}). This explains why the observed long term variability at 850 $\mu$m is not apparent in the $W1$ and $W2$ light curves of the source.

Within the sample of 126 matched source candidates, ten have {\it WISE} and JCMT emission that probably arise from different sources.   Of the remaining 116 sources, 57  have unreliable {\it WISE} photometry due to crowding, bright neighbours, or being spurious detections. 

These brightness and positional matching criteria leads to a selection of 59 objects where JCMT fluxes and {\it WISE} emission are likely produced by the same source. 

\subsection{Earlier sub-mm data}

Of the 59 objects in this final sample, 35 have archival fluxes from the JCMT Gould Belt Survey  \citep[GBS;][]{2007Ward}, as  published in the variability analysis of \citet{2017Mairs}.  These Gould Belt Survey observations were taken between 2012-2014, which allows us to investigate mid-IR to sub-mm variability over timescales of up to 6 years. The flux for objects with a positive crossmatch were calibrated using the conversion factors determined by \citet{2017Mairs}. Following the analysis of \citet{2017Mairs} we only used the mean flux of the additional GBS epochs (we found between 3 to 7 additional epochs for objects with a positive crossmatch).    

\subsection{Defining a time window}

Finally, because we are interested in finding correlated variability between the mid-IR and sub-mm flux, we visually inspect the light curves of the 59 objects in our sample to define time windows that provide the best chance to find such correlation. Defining a contemporaneous dataset is challenging given the gaps between the GBS and Transient surveys data as well as the gap between {\it WISE} and {\it NEOWISE} observations. The window is not the same for every object and is defined mainly on the behaviour of the sub-mm flux and by maximising the number of contemporaneous observations between both surveys. 

For example, source 17 in NGC1333 \citep[ASR 32,][]{1994Aspin} shows an apparent linear decay for MJD$>57300$~d at 850 $\mu$m (see Fig.~\ref{fig:n1333_17}). Selecting data with MJD$>57300$~d ensures that we are analysing the main region of interest from the sub-mm flux as well as selecting a large number of nearly contemporaneous {\it WISE} observations. Setting this lower time limit for the source meant that we did not include the GBS data nor the first epochs of the JCMT Transient survey when calculating the statistical measures of variability. Even though the selection of the window is made from the apparent variability of the sub-mm flux, this selection does not imply that that the variability is statistically significant.  The apparent linear decay in the sub-mm flux of ASR32 (Fig. \ref{fig:n1333_17}), is not found to be statically significant in our analysis ($|S/\Delta S| <3$, see Section \ref{sec:statvar}). 

\subsection{Final source catalogue}

Columns 1 to 8 in Table \ref{tab:allvar} show the JCMT Transient survey region, source number, right ascension, declination, JCMT designation, most common name from the literature (taken from the SIMBAD database) , YSO class and reference for the YSO classification, for the 59 sources. In cases where we do not find any information in the literature to classify the YSOs, the classification arises from SED inspection and is marked with a ? sign. Columns 9 to 14 present the average $W2$ magnitude, $\Delta W2 (90\%)$ (or the 90th minus 10th percentile in magnitude), $\Delta W2$ (using all of the available data in the light curve), and the measurements of statistical variability SD/SD$_{\rm fid}$ and $S/\Delta S$  (in flux units and defined later in Section \ref{sec:statvar}). Column 14 shows the number of {\it WISE} points used in the analysis of variability. Columns 15 to 18 show data obtained from the JCMT. These correspond to the mean peak flux over the analysed epochs,  SD/SD$_{\rm fid}$, $S/\Delta S$ and the number of epochs used in the analysis of variability.

To be consistent with the analysis of \citet{2018Johnstone}, the statistical measurements of stochastic and secular variability (to be defined below) for the mid-IR data are determined using {\it WISE} fluxes, with conversions using zero magnitude flux densities of F$_{\nu,0}=309.54$ Jy and  F$_{\nu,0}=171.787$ Jy respectively \citep{2012Cutri}.

\section{Statistical search for variability}\label{sec:statvar}

\begin{figure*}
	\resizebox{0.97\columnwidth}{!}{\includegraphics{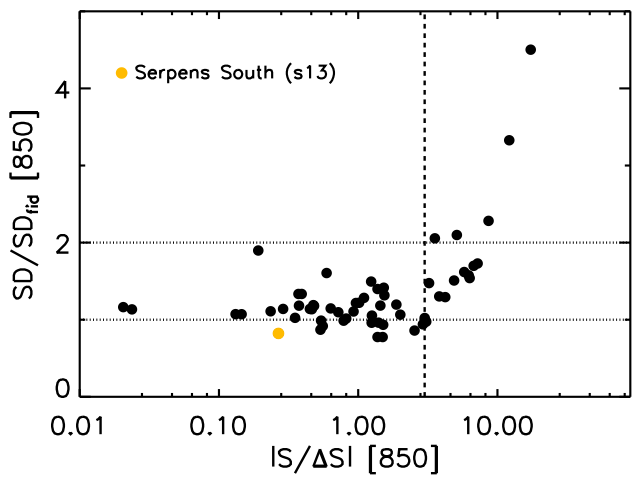}}
	\hspace{0.5cm}
	\resizebox{0.99\columnwidth}{!}{\includegraphics{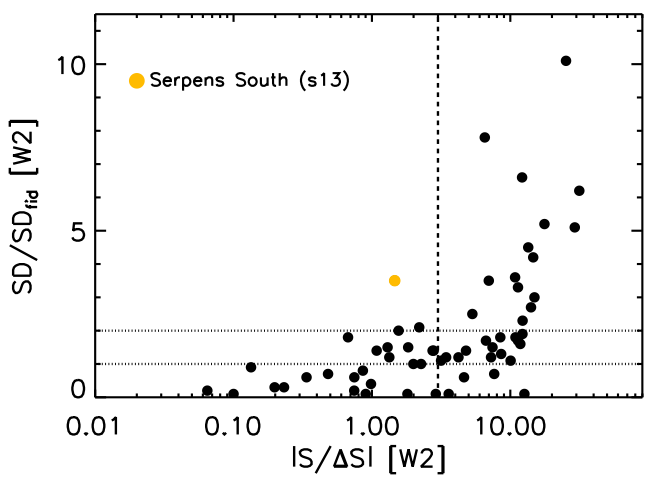}}
    \caption{SD/SD$_{{\rm fid}}$ vs $|S/\Delta S|$ for JCMT (left) and {\it WISE} $W2$ (right) data. In the plots, horizontal dotted lines mark the levels of SD/SD$_{{\rm fid}}$ equal to unity and SD/SD$_{{\rm fid}}=2$. Objects found above the latter level are found to be stochastically variable objects. The vertical dashed lines mark $|S/\Delta S|=3$. Objects to the right of this line are candidates to display secular variability. The location of source 13 in Serpens South (see main text) is indicated by the orange solid circle. This object is discussed in more detail in Appendix \ref{sec:ind_sto}.}
    \label{fig:stat1}
\end{figure*}

Using the JCMT and {\it WISE} data for the sample of 59 protostars selected  above, we search for signs of variability following a similar analysis done by \citet{2018Johnstone}.  This discussion presents a general description of the variability of the whole sample. Discussions of individual objects of interest is presented in Appendix \ref{sec:ind}.

In the analysis of the first 12 epochs of the JCMT Transient survey, \citet{2018Johnstone} searched for signs of stochastic and secular variability over the 8 regions studied by the survey. First, \citet{2018Johnstone} determined the standard deviation of the observed JCMT light curves and compared them to a fiducial model of the expected uncertainty for each source, SD$_{\rm fid}$\citep[given by equation 1 in][]{2018Johnstone}. The comparison of the standard deviation against the fiducial model,  SD/SD$_{{\rm fid}}$, provides an indication of the stochastic variable behaviour of the source. In the search for secular variability, \citet{2018Johnstone} perform a linear fit to the JCMT brightness measurements. Comparison of the slope against the uncertainty in the slope measurement is used as a signpost of the object showing secular variability, i.e. a monotonic rise (or fading) with time. \citet{2018Johnstone} found 5 statistical outliers that show long term brightness changes across the early data of the JCMT Transient Survey. One source with a standard deviation of brightness significantly larger than the expected level, the protostar EC53 \citep[][see also later in this work]{2017Yoo}, was also among objects with signatures of secular variability. For a given source long-term trends will increase the stochasticity of the object.

We perform a statistical investigation in search for variability in our sample, following the analysis of \citet{2018Johnstone}. For our sample of roughly 30 submillimetre epochs per source we  determine SD/SD$_{\rm fid}$ using the JCMT  and {\it WISE} observations that are within the time limits defined for each source (see above). For the {\it WISE} data we also determine the standard deviation of the $W1$ and $W2$ fluxes and define the expected uncertainty, SD$_{\rm fid}$, simply as the average error across all epochs.

To search for secular variability in the {\it WISE}  and JCMT data we used IDL {\sc{linfit}} to perform a least-squares linear fit to the $W1$, $W2$ and 850 $\mu$m fluxes \citep[see also section 4 of ][]{2018Johnstone}, 

\begin{equation}
f(t)=f_{0}(1+S(t-t_{0})),
\end{equation} 

\noindent with $S$, the slope of the fit, $f_{0}$ and $t_{0}$ the flux and MJD of the first {\it WISE} or JCMT epochs, respectively. The IDL procedure is based on the {\sc{fit}} and {\sc{gammq}} routines described in \citet{1989Press}. IDL {\sc{linfit}} also returns the uncertainty in $S$, $\Delta S$. The ratio $S/\Delta$ S provides an indication of how good the observations can be described by a linear increase(decrease) with time, with  $|S/\Delta S|>3$ defined as where the linear variability is statistically significant.  In the following analysis we  focus mainly on the results obtained from the $W2$ band, since many objects have faint $W1$ photometry near the sensitivity limit of the individual observations.

\begin{figure}
\resizebox{1\columnwidth}{!}{\includegraphics{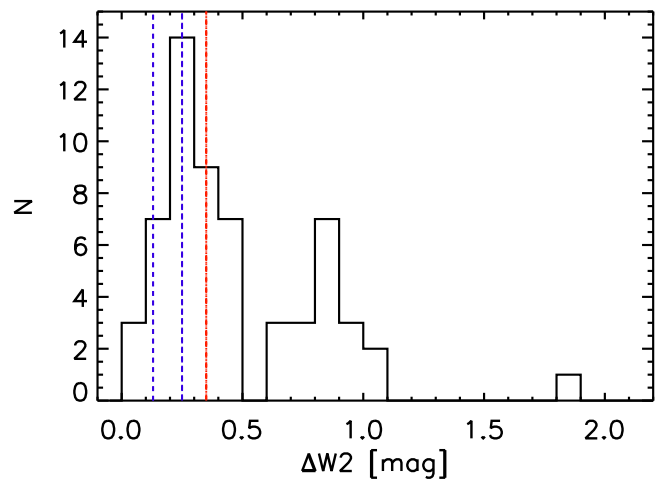}}
\caption{Histogram of $W2$ amplitudes (defined as the 90th minus 10th percentile in magnitude) for 59 sources in our sample. The blue dashed line marks the range of median amplitude for class I YSOs in different star forming regions by YSOVAR  \citep[see e.g.][]{2014Gunther, 2018Wolk}, while the red dot--dashed line marks the median amplitudes for our sample of YSOs.}
    \label{fig:w2hist}
\end{figure}

Fig. \ref{fig:stat1} shows the comparison of SD/SD$_{\rm fid}$ vs $|S/\Delta S|$ for the 850 $\mu$m data for the 59 sources in our sample. The results are similar to those from the analysis of \citet{2018Johnstone}, where the majority of YSOs in our sample that display variability in JCMT data are showing secular changes. From the sample, 16 sources show $|S/\Delta S|\geq3$.  Only 5 sources show  a high degree of stochastic variability at $850 \mu$m (defined as  objects SD/SD$_{\rm fid} \geq 2$). Similar to what was found by \citet{2018Johnstone} , Source 2 in the Serpens Main region \citep[EC53, see e.g.][]{2017Yoo} shows the largest standard deviation over the fiducial level in our sample. This is driven by quasi-periodic eruptions every 520-570 days \citep{2012Hodapp,2017Yoo}. The source also shows long-term trends that might help to increase SD/SD$_{\rm fid}$. In the remaining four cases we do not observe a similar periodic behaviour. The variability of these sources, however, remain dominated by long timescale brightness changes and this secular behaviour is responsible for the enhanced  SD/SD$_{\rm fid}$.

Fig. \ref{fig:stat1} also shows the comparison of SD/SD$_{\rm fid}$ vs $|S/\Delta S|$ for the {\it WISE} W2 data. Once again the variability is dominated by long-term changes in the brightness of the sources, with 33 out of 59 objects showing signs of secular variability. In contrast with the JCMT data, a large number of YSOs in the sample show stochastic variability in their {\it WISE} $W2$ light curves, with 18 YSOs showing SD/SD$_{\rm fid} \geq 2$. The ability to detect the stochastic variability in the {\it WISE} data is likely due to the higher signal to noise ratios at these wavelengths compared with the JCMT Transient Survey (see also Section \ref{sec:wisevar}). 

The fact that the majority of the sample shows variability in the mid-IR data of {\it WISE} is not surprising and agrees with the results from the YSOVAR campaign, where $\simeq80 \%$ of protostars are found to show variability at these wavelengths \citep[see e.g.][]{2018Wolk}. However, the sample of 59 YSOs shows median $W2$ amplitudes \citep[defined as the 90th minus 10th percentile in magnitude to remove outlying points, following the similar approach of][]{2018Wolk} of 0.35 mag (see Fig. \ref{fig:w2hist}), which is larger than the median amplitudes of 0.13 and 0.25 mag at 4.5 $\mu$m for class I YSOs in star forming regions studied by the YSOVAR team \citep{2014Gunther, 2015Poppenhaeger, 2015Rebull, 2015Wolk, 2018Wolk}. Many objects in our sample are located around the median amplitudes observed by YSOVAR, but Fig. \ref{fig:w2hist} also shows a large number of sources that have larger amplitudes and might correspond to a population of extreme variable stars. The latter probably indicates that the requirement of a sub-mm detection selects YSOs at earlier evolutionary stages.

In Fig. \ref{fig:stat1} only one object stands out clearly as a possibly purely stochastic variable star (without any apparent secular changes) in the {\it WISE} data: source 13 in the Serpens South region \citep[2MASS J18300101$-$0206082;][]{2015Dunham}. There are no objects in the purely stochastic area in the JCMT data. The variability of 2MASS J18300101$-$0206082 is of low amplitude and may be driven by variable extinction. These changes are not observed at sub-mm wavelengths. A more detailed discussion on this object is presented in Appendix \ref{sec:ind_sto}. 

\section{Secular Variability}\label{sec:wisevar}

In this section we describe the correlations between the mid-IR and sub-mm variability.  The previous analysis shows that the measured flux variability between epochs is dominated by long-term secular changes and not random stochastic variability, especially in the sub-mm. Removing the long-term trends could allow the study of correlated short-term stochastic variability. However, this is challenging due to the lower cadence of {\it WISE} compared with the JCMT data, and thus the fewer data points available for analysis. In addition brightness changes on short timescales are likely to be small at 3.6 and 4.5 $\mu$m \citep[see e.g.,][]{2014Cody}. In the following we will focus only on the secular changes observed in the data from the JCMT and {\it WISE} surveys.

\begin{figure*}
	\resizebox{2\columnwidth}{!}{\includegraphics{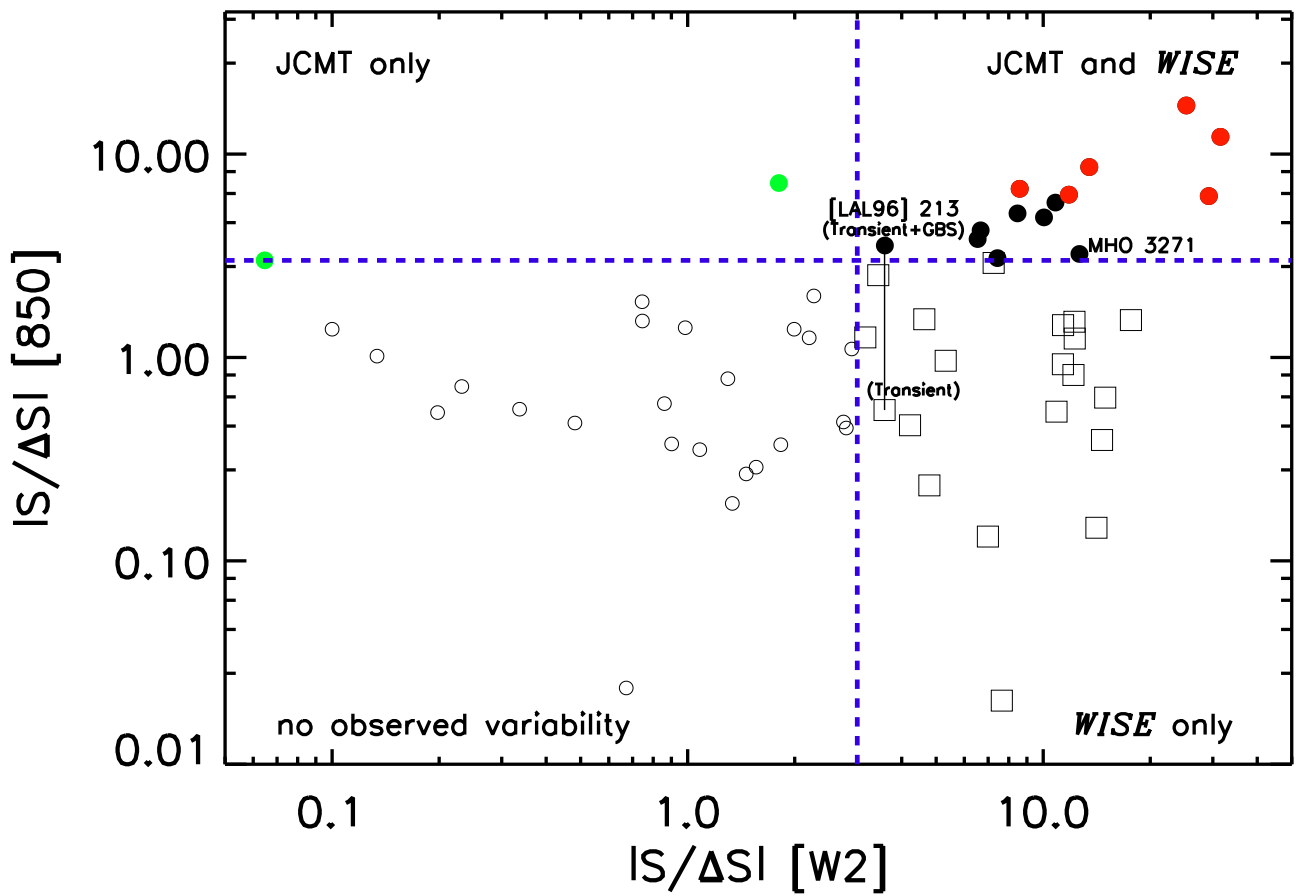}}\\
	    \caption{$|S/\Delta S|$ (JCMT) vs $|S/\Delta S|$ ({\it WISE} $W2$) for the 59 targets in our sample. The blue dotted lines mark $|S/\Delta S|=3$, objects with values larger than this limit are strong variable candidates in both surveys. Depending on where objects locate we can define 4 regions of variability. In the bottom left quadrant we find non variable objects (open circles), the upper left quadrant shows objects with variability only in the JCMT Transient survey (green circles), while the bottom right quadrant defines the region where we only observe variability at $W2$ (open squares). Finally objects that are found to be variable in both surveys are located in the upper right quadrant (solid circles). In the latter region objects with the largest variability in both surveys $|S/\Delta S| \geq6$  are marked by the solid red circles. In the figure we mark the location of YSO [LAL96] 213, an object that moves between different regions when using additional data arising from the JCMT Gould Belt survey (see Section \ref{sec:wisevar}), and YSO MHO 3271, an object that shows uncorrelated secular variability.}
    \label{fig:ex3}
\end{figure*}

\begin{figure}
	\resizebox{0.98\columnwidth}{!}{\includegraphics{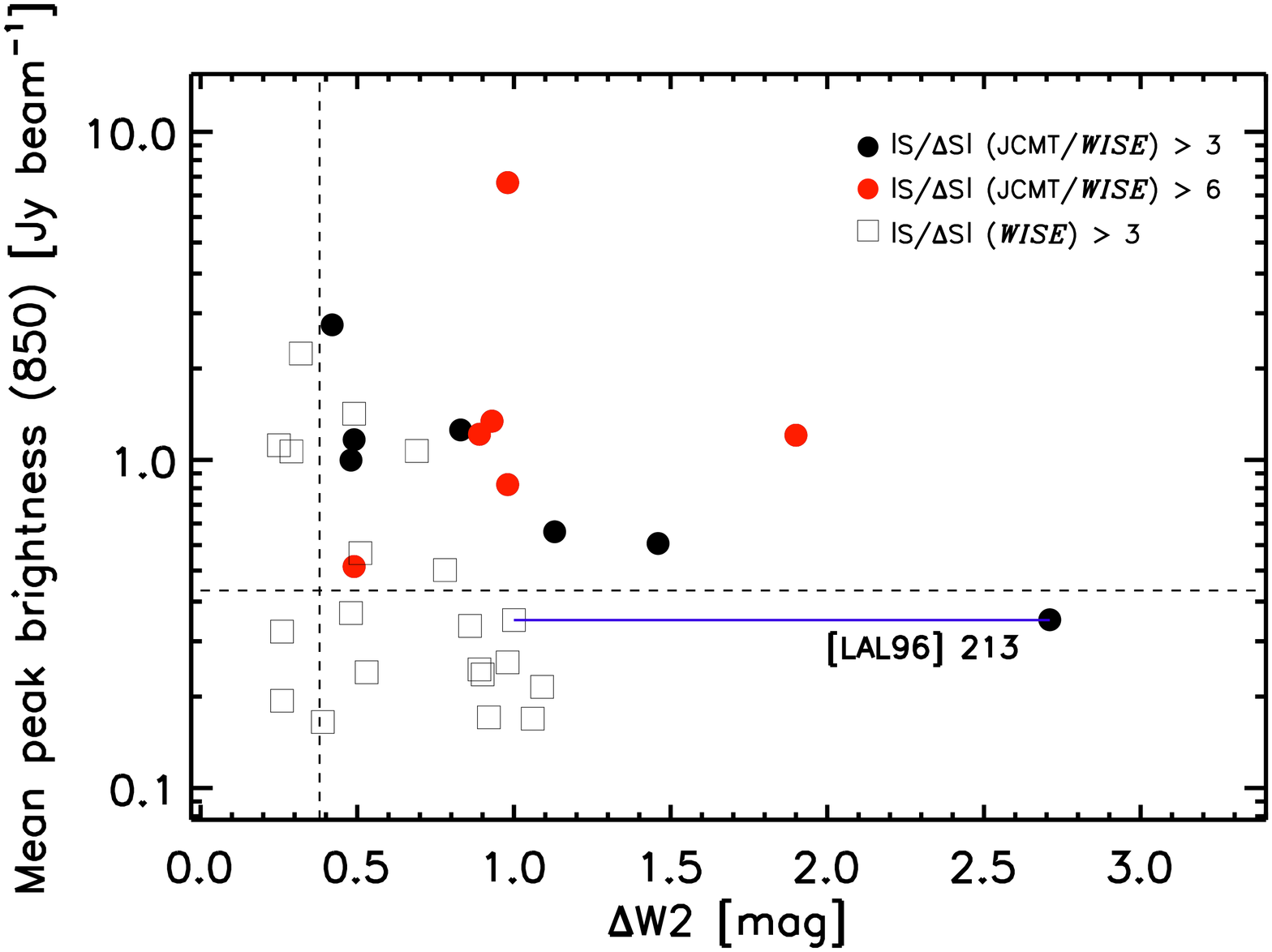}}\\
	\resizebox{0.98\columnwidth}{!}{\includegraphics{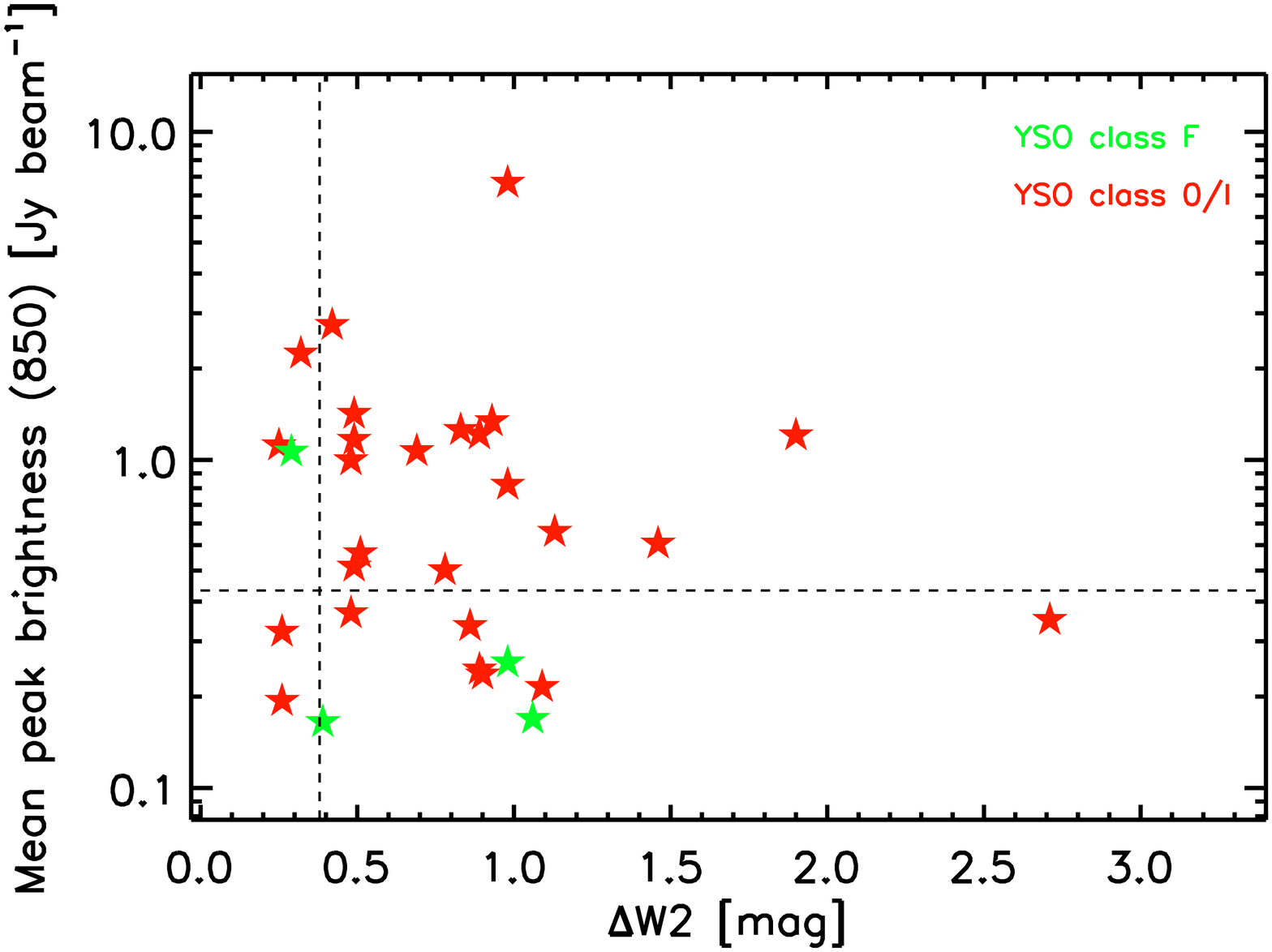}}\\
	\resizebox{0.95\columnwidth}{!}{\includegraphics{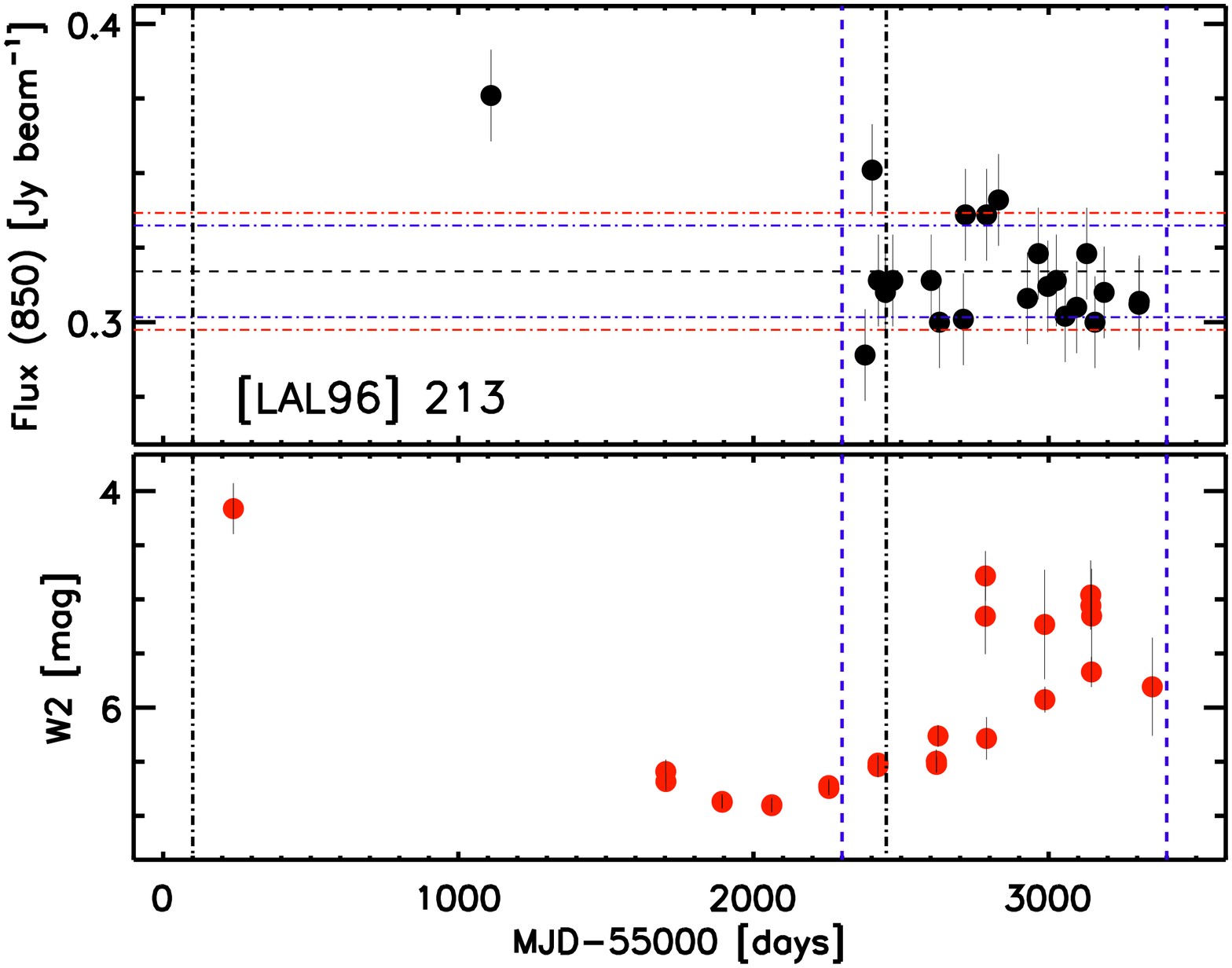}}
		    \caption{(Top) Mean peak brightness at 850 $\mu$m vs $\Delta W2$ for sources that are found to be variable only in {\it WISE} (open squares) and sources that are variable in both surveys (red and black circles). (Middle) Same as previous plot, but this time dividing objects according to YSO class, with class 0/I sources shown as red stars, while older flat-spectrum sources are shown as green stars. In the top and middle figures, dashed lines mark the limits of 0.38 mag (vertical line) and 0.4 Jy beam$^{-1}$ (horizontal line) discussed in the text. (Bottom) 850 $\mu$m and {\it WISE} $W2$ light curves for [LAL96] 213. In the plot, the black dot-dashed lines mark the time between the GBS and the first few epochs of the JCMT Transient survey, while the dashed blue lines encompass only the data from the JCMT Transient survey. The variability of the YSO was studied during these two periods of time and it appears to show that variability at $850 \mu$m is only detected when the mid-IR variability surpasses 3 magnitudes (see main text).}
    \label{fig:class}
\end{figure}

Figure \ref{fig:ex3} shows the comparison of the {\it WISE} $W2$ vs JCMT  statistical measure of secular variability ($S/\Delta S$) for the 59 sources in our sample. Sources with values of $|S/\Delta S|\geq3$ are strong candidates for variability in both JCMT and {\it WISE}. Using these limits we can define four quadrants of variability in Fig. \ref{fig:ex3}:  objects with no observed variability (bottom left), with apparent variability only in JCMT data (top left), with apparent variability only in {\it WISE} data (bottom right) and with secular variability in both surveys (top right).

Source 11 in Serpens South \citep[MHO 3271,][]{2015Zhang} is the only object in the top-right quadrant of Fig. \ref{fig:ex3} for which the secular brightness change is anti-correlated. The object gets brighter in the sub-mm (positive value of S/$\Delta$S) and fainter in the mid-IR (thus, negative values of S/$\Delta$S). However the value for {\it WISE} is measured from only a few epochs (see discussion in Appendix \ref{sec:ss11}) and it is difficult to conclude that the observed lack of correlation is real. This object also falls very close to the $S/\Delta S=3$ line for the JCMT data. Given this, the source is not included in any further analysis.


In the following we will discuss three of the four quadrants of variability. The discussion of the most interesting quadrant, where objects show secular variability over the two surveys, is done in the next section of this paper.

\begin{itemize}
\item[] {\it No observed variability:} In 24 YSOs significant variability is not observed in either survey. The sources located in this region will not be discussed any further.

\item[] {\it JCMT only:} In the top left corner of Fig. \ref{fig:ex3}, two objects, source 5 in NGC2068 \citep[HBC 502;][]{1988Herbig} and source 0 in NGC1333 \citep[IRAS4A;][]{1987Jennings}, show significant secular variability in JCMT but not in {\it WISE}.  For IRAS4A this is likely explained by the low number of reliable {\it WISE} data points.  On the other hand, the lightcurves of HBC 502 are well sampled.  The variability in JCMT but not WISE may be explained if the mid-IR and sub-mm emission do not arise from the same source, for example if the IR emission is dominated by outflows. Figures and a further discussion on these two objects are presented in Section \ref{sec:ind_jcmt}. 
 
\item[] {\it WISE variability:} The bottom right corner of Fig. \ref{fig:ex3} shows a large number of objects with significant {\it WISE} variability but with no apparent correlation in the JCMT Transient survey data. Both the mean peak brightness at 850 $\mu$m and the amplitude of the {\it WISE} $W2$ variability appear to play a role in determining whether objects display variability over the two surveys. 

The top plot of Fig. \ref{fig:class} shows the mean peak brightness at 850 $\mu$m (JCMT) versus $W2$ amplitude (using all the points in the lightcurves) for objects with $|S/\Delta S| ($W2$) \geq 3$. All of the objects with mean peak 850 $\mu$m brightness fainter than 0.4 Jy beam$^{-1}$, except for [LAL96] 213, show $\Delta W2<1.5$~mag and are not identified as variables in the JCMT data. Fig. \ref{fig:class} also shows that objects with {\it WISE} variability but with $\Delta W2 \leq 0.38$~mag are also not found to be variable in JCMT data, independent of the mean brightness of the source at 850 $\mu$m.

The exception, YSO [LAL96] 213, shows that below a mean peak brightness of 0.4 Jy beam$^{-1}$, variability with amplitude larger than $\simeq$2.5 in {\it WISE} is required to detect any variability at 850 $\mu$m. The bottom plot of Fig. \ref{fig:class} shows the mid-IR and sub-mm light curve of [LAL96] 213. Considering only the data from the JCMT Transient survey, the object is not found to be variable in the sub-mm, despite showing $\Delta W2\sim 1$-$1.5$ mag in the same period of time.  When the earlier epochs are included in this analysis, the $\Delta W2$ surpasses $\simeq2.5$ magnitudes, and the object moves towards the region with mid-IR and sub-mm variability (see Fig. \ref{fig:ex3}).

The brightness and amplitude limits discussed above prevent us from studying any correlation between the mid-IR and sub-mm variability in more evolved YSOs, which are typically faint in the sub-mm. In Fig. \ref{fig:class} (middle panel) we can see that flat spectrum sources are located below either the brightness limit of 0.4 Jy beam$^{-1}$ at 850 $\mu$m or the amplitude limit of 0.38 mag. None of the six class II YSOs within our sample of 59 sources is found to be a secular variable in {\it WISE}. The analysis of any correlations between mid-IR and sub-mm variability in this paper is therefore restricted to objects that are Class I or earlier.

Four objects are variable in {\it WISE} but not in JCMT data and are located above the brightness and amplitude limits discussed above (source 18 in NGC1333, source 6 in OMC2/3, source 4 in NGC2068 and source 35 in NGC1333). This can be explained by several effects, such as extinction or the sub-mm and mid-IR fluxes not arising from the same source.  Each of these sources is described in more detail in Section \ref{sec:ind_wise}.

\end{itemize}

\begin{figure}
	\resizebox{0.9\columnwidth}{!}{\includegraphics{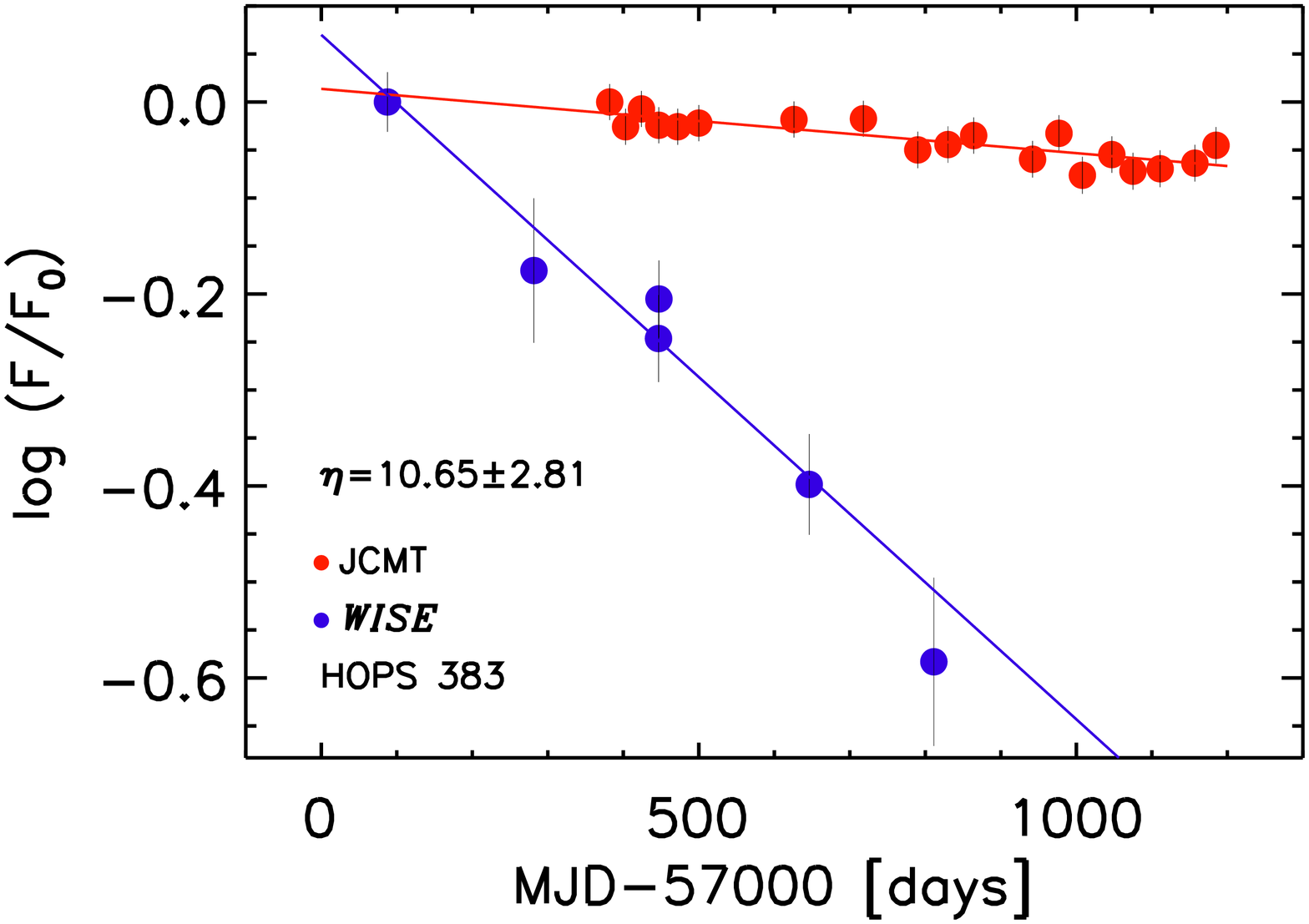}}\\
	\resizebox{0.9\columnwidth}{!}{\includegraphics{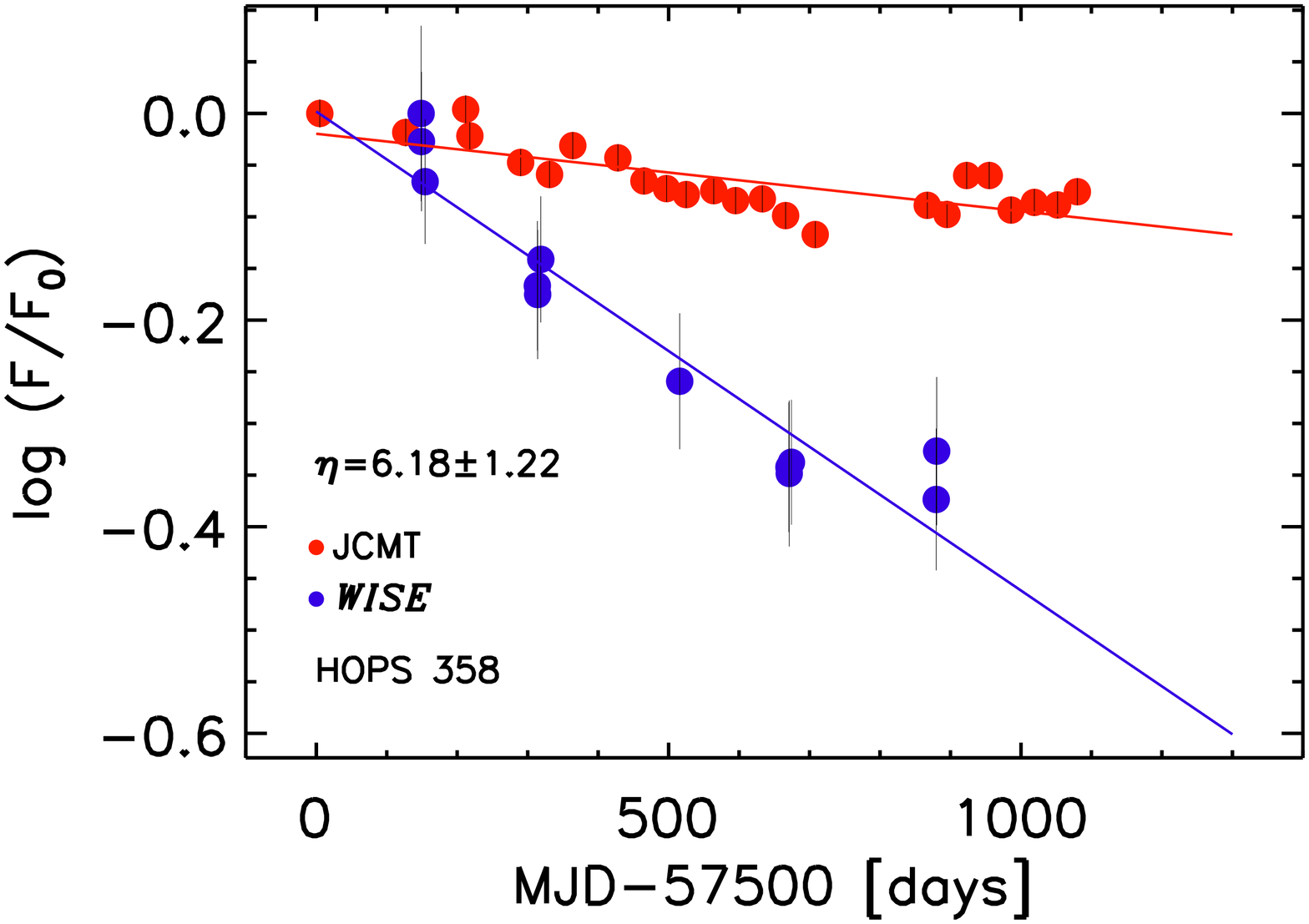}}\\
	\resizebox{0.9\columnwidth}{!}{\includegraphics{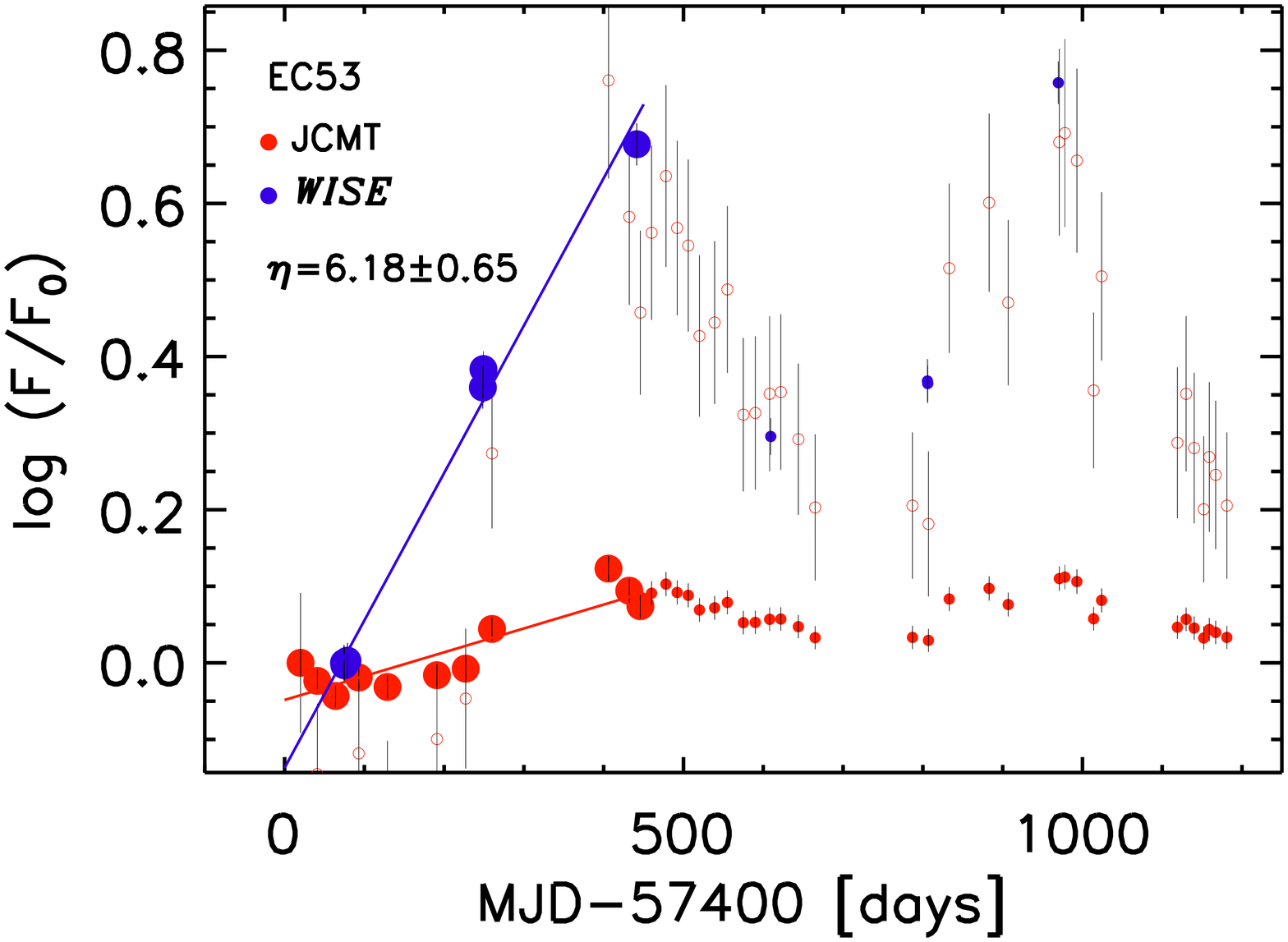}}
    \caption{Examples of the fits to $\log$ Flux for source 52 in OMC2/3 (HOPS 383, top), source 1 in NGC2068 (HOPS 358, middle) and source 2 in Serpens Main (EC53, bottom). The fits to the {\it WISE} $W2$ and JCMT data are shown as solid blue and red lines respectively. In all panels the values of $\eta$ obtained for each individual case are shown for reference. In the bottom panel large solid circles marks the data points that were used in the fits for EC53, while small solid circles show the remaining observations. Red open circles show the result of scaling the sub-mm flux using $\eta$ obtained for EC53.}
    \label{fig:logf}
\end{figure}

\section{Mid-IR to sub-mm continuum variability}\label{sec:wjvar}

The analysis of the previous section made clear that out of 59 sources with bright enough sub-mm and accurate enough {\it WISE} data, 13 sources show signs of secular variability in both the {\it WISE} and JCMT Transient survey data (objects falling in the top right corner of Fig. \ref{fig:ex3}). These represent $22\%$ of the clean {\it WISE} sample. Columns 1 to 4 in Table \ref{tab:allsec} show the name from the literature, YSO class, $(S/\Delta S)$ at 850 $\mu$m and $W2$ for the 13 sources. Following the analysis of saturated {\it WISE} sources from Appendix \ref{app:sat}, YSO V1647 Ori is also included in the list. All YSOs presented in Table \ref{tab:allsec} are classified as Class 0 or I sources. This could suggest that long-term variability across the mid-IR and sub-mm is a property of YSOs at early evolutionary stages.

\begin{table}
	\centering
	\caption{The 14 sources with variability in both {\it WISE} $W2$ and the JCMT Transient survey.}
	\label{tab:allsec}
\resizebox{\columnwidth}{!}{
   	\begin{tabular}{lccc} 
		\hline
		Name & YSO class & $S/\Delta S$ (JCMT) & $S/\Delta S$ ({\it WISE}) \\
		\hline
	     CAZ2013 IC348MMS1 & I & 6.3 & 11.8 \\
		SSTc2d J032904.1$+$311447 & I &  6.8 &  8.6 \\
		$[$LAL96$]$ 213 & 0 &  $-$3.6 & $-$3.6 \\
		HOPS 317 & 0 & 3.1 &  7.4 \\
		HOPS 358 & 0 & $-$12.2 & $-$31.5 \\
		 HOPS 373 & 0 & $-$4.9 &  $-$10.1 \\
		HOPS 323 & I &  $-$4.2 & $-$6.7 \\
	     HOPS 324 & I &  3.8 & 6.5 \\
		 $[$CHS2001$]$ 8787 & I &   5.1 &   8.5 \\
		HOPS 383 & 0 & $-$5.8 & $-$10.8 \\
		Serpens SMM1 & 0 & 8.6 & 13.5 \\
	     EC53 & I & 17.3 & 25.3 \\
		Serpens SMM10IR & I &  6.2 & 29.2 \\
		 V1647 Ori$^{\dagger}$ & I & -9.7 &-12.7\\ 
		\hline
		\multicolumn{4}{l}{$\dagger$ This YSO is included from the analysis of Appendix \ref{app:sat}.}
	\end{tabular}}
\end{table}  

\subsection{Correlated Variability}

The primary goal of this paper is to determine whether and how sub-mm and mid-IR emission variability is correlated. Past efforts have usually been limited to individual objects with outburst and pre- or post-outburst SEDs \citep[e.g.][]{2007Kospal,Juhasz2012,2013kospal,2015Safron}.

According to the arguments presented in Section \ref{ssec:jcmt_desc}, and as indicated by the radiative transfer simulations included in \citet{2013Scholz}, for deeply embedded sources the 850 $\mu$m emission traces the temperature changes in the envelope, while the mid-IR flux traces the emission from the protostar and inner disc, which is expected to follow more closely to the accretion luminosity.

\begin{figure*}
	\resizebox{1.5\columnwidth}{!}{\includegraphics{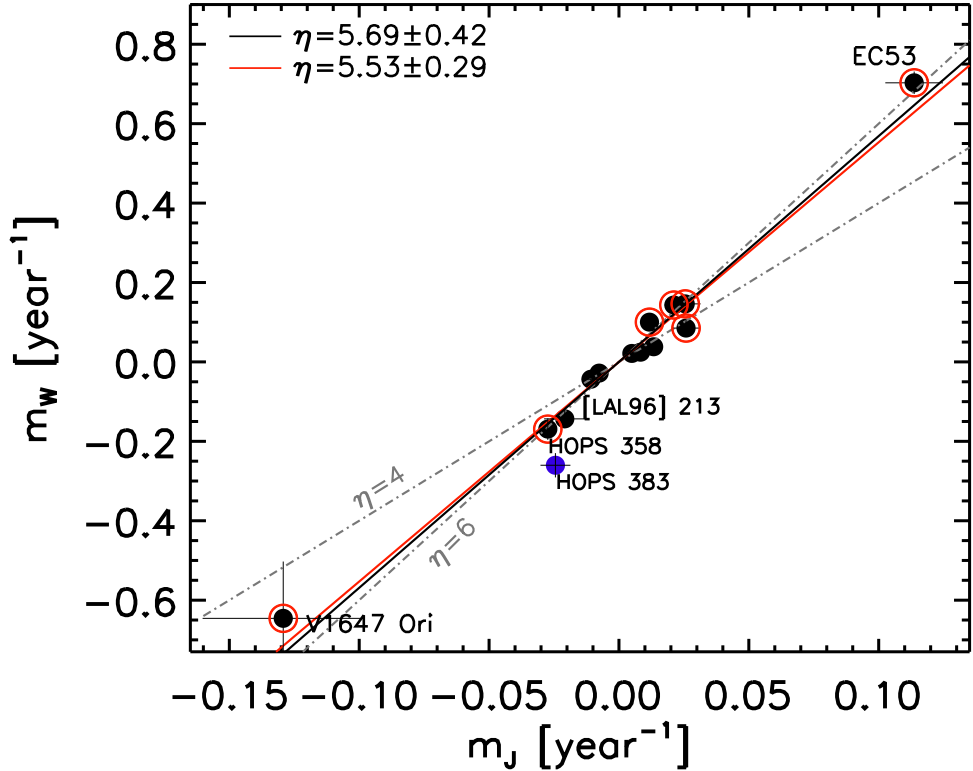}}
    \caption{$m_{w}$ vs $m{j}$ for sources in our sample that show variability in both {\it WISE} and the JCMT surveys. Red circles mark objects with large variability in both surveys ($|S/\Delta S| \geq 6$). We show the least-squares fits using the OLS bisector expression in table 1 of \citet{1990Isobe} for all of the sources (black solid line) and only using sources with large variability (red solid line). In the figure, the grey dash-dotted lines show the fits using $\eta=4$ and $\eta=6$, representing the range of values expected from Section \ref{ssec:initjcmt}.} 
    \label{fig:slope}
\end{figure*}

Assuming that during the time of contemporaneous observations in both surveys the measured $\log$(Flux) follows a linear function in time, then

\begin{eqnarray}
\log \left( \frac{F_{x}(t)}{F_{x,0}} \right)=m_{x}(t-t_{0}),
\end{eqnarray}

\noindent where the subscript x stands for either {\it WISE} (w) or JCMT (j) observations, $F_{x}$ is the measured flux and  $m_{x}$ is the slope of the linear relation. Further, assuming that the changes in flux are correlated such that

 \begin{eqnarray}
\log \left( \frac{F_{w}(t)}{F_{w,0}} \right)= \eta \log \left( \frac{F_{j}(t)}{F_{j,0}} \right),
\end{eqnarray}

\noindent then for sources observed over the same time intervals

\begin{eqnarray}
m_{w}=\eta m_{j}.
\end{eqnarray}

For both {\it WISE} and JCMT, linear models are fitted to the 14 sources of Table \ref{tab:allsec}. Figure \ref{fig:logf} shows examples of the fits for 850 $\mu$m and {\it WISE} $W2$ fluxes for targets HOPS 383, HOPS 358 and EC 53. These fits differ to those in Section \ref{sec:statvar}, since here the relationship defined by $\eta$ holds over $\log$ Flux space, while slopes S derived in Section \ref{sec:statvar}  are estimated directly from the mid-IR and sub-mm fluxes. 

The  YSO EC53 lightcurve is treated differently than the other sources because the source undergoes quasi-periodic eruptions.  The first burst in our time series has best coverage in the combined WISE-JCMT dataset and is adopted for our analysis. Scaling the sub-mm flux using the factor ($\eta$) obtained from this fit shows a good agreement with the {\it WISE} data over the whole light curve (see Fig. \ref{fig:logf}) and agrees well with the scaling factor obtained using a more detailed multi-wavelength, periodogram-based fit of EC53 (Y.-H. Lee et al., in prep).

The values of $m_{w}$ and $m_{j}$ are related to each other by $\eta=5.53\pm0.29$ (Fig.~\ref{fig:slope}), based on a least-squares fit to objects  with the most significant variability in both surveys ($|S/\Delta S| \geq 6$).  The fit uses the ``OLS bisector'' expression in Table 1 of \citet{1990Isobe}.  This value is adopted for all further analysis.  
Across the sample of fourteen sources, including less significant variables, the fit would instead be $\eta=5.69\pm0.42$.

Fig. \ref{fig:slope} shows that there is some scatter around the $\eta=5.53$ line.  One object, HOPS 383, falls well below this line, indicating that for this source the mid-IR emission falls more sharply than expected from the sub-mm and the $\eta=5.53$ relationship (see Fig. \ref{fig:logf}). The decline in the mid-IR may either be caused by extinction or by a sharp decline in the disk luminosity.

\subsection{Sub-mm and Dust temperature Response to Accretion Luminosity}

\citet{2019Macfarlane_b} performed radiative transfer modelling of eruptive YSOs exploring a wide range in outburst luminosities and properties of these systems. Figure 3 in \citet{2019Macfarlane_b} shows the SED variation for different outburst luminosities in one of their models. Inspection of this figure shows that at sub-mm wavelengths we see only a small change in the flux. At around 100 $\mu$m the  flux is directly proportional to the luminosity.  At shorter wavelengths we observe a much more complicated relationship, which reflects a dependence on the structure of the envelope of the system.

The specific response of the sub-mm brightness to changes in the envelope dust temperature is estimated by noting that the emission scales directly with the Planck function. Thus:
\begin{eqnarray}
F_{850}(T_d) \propto (e^{(-T_\nu/T_d)} -1)^{-1},
\end{eqnarray}
where T$_\nu$ = h $\nu$/k = 17 K at 850 microns. Taking the derivative with respect to T$_d$ yields
\begin{eqnarray}
\frac{d ln F}{d ln T_d} = \frac{(T_\nu/T_d) e^{(T_\nu/T_d)}}{(e^{(T_\nu/T_d)} - 1)}
\end{eqnarray}
For T$_d$=20 K, a typical dust temperature in the outer envelope where the bulk of the sub-mm emission arises, the 850 $\mu$m flux varies as F$_{850} \propto T_{d}^{1.5}$, a somewhat stronger than linear response due to the fact that at such low temperatures the emission at 850 $\mu$m is not yet fully on the Rayleigh-Jeans tail.

\citet{larson69} showed that, in the optically thin limit and a fixed source luminosity, the envelope dust temperature radial profile will be flatter than $T_d \propto r^{-1/2}$ when the dust opacity does not mimic a grey body but instead is more emissive at higher frequencies. Quantitatively, if the dust opacity follows a power-law, $\kappa \propto \nu^{\beta_{\rm em}}$, across the frequencies at which it primarily emits, then the dust temperature radial profile will be $T_d \propto r^{-2/(4+\beta_{\rm em})}$.

When the luminosity of the central source is changing, it becomes more complicated to determine the dust temperature temporal profile at a fixed position in the envelope. The equilibrium dust temperature in the outer, optically thin envelope is set by balancing absorption and emission, but in this case the fractional absorption of incident energy is not constant. Thus, both the strength of the emission and absorption depend on the dust opacity law at the frequencies of emission and absorption of photons by the dust \citep{2015Ryden}, with the absorption dependent on the specific shape of the incident radiation field  (i.e.\ the effective temperature or hardness of the radiation field). If the temperature of the radiation field remains fixed while the luminosity varies, then the dust temperature response will depend only on the dust opacity-law for emission. In this case, $T_d \propto L^{1/(4+\beta_{\rm em})}$. On the other hand, if the source luminosity follows a blackbody formulation with $L \propto T_{\rm ph}^4$, then balancing absorption and emission requires that $T_d \propto L^{(1+\beta_{\rm abs}/4)/(4 + \beta_{\rm em})}$.  For a grey opacity at the frequencies where the source luminosity is absorbed, $\beta_{\rm abs} \sim 0$, the relation reduces to the previous dust temperature relation, since in this case the shape of the incident radiation field does not matter. 

For deeply embedded protostars, the temperature at the effective photosphere is expected to vary only slightly with accretion luminosity \citep{1998Hartmann}, thus we expect $T_d \propto L^{1/4+\beta_{\rm em}}$. In this case, assuming $T_d \sim 20$\,K,  the 850 $\mu$m brightness will vary as
 \begin{eqnarray}
F_{850} \propto L^{1.5/(4+\beta_{em})}
\end{eqnarray}
Consequently, the observed scaling between the sub-mm and mid-IR reduces to
 \begin{eqnarray}
F_{IR} \propto L^{8.3/(4+\beta_{em})}.
\end{eqnarray}
Finally, if we assume $\beta_{\rm em} \sim 1.5$, then F$_{850} \propto$ L$^{0.27}$ and F$_{\rm IR} \propto$ L$^{1.5}$.

\subsection{Numerical SED Models of Variable Deeply Embedded Protostars}

For envelopes around protostars, variations in the source luminosity produce changes to both the radius and the temperature at the effective photosphere of the envelope, defined as the location where the bulk of the radiation energy being emitted starts to become optically thin \citep[for example, see][]{2013Johnstone}.  Since the shape of the radiation changes as the source luminosity increases, detailed radiative transfer models are essential to capture the nuances in the resulting SED as a function of changing luminosity and thus determine the expected $\eta$.  As an example, \citet{2020Baek} used 2-D and 3-D radiative transfer models to fit the SED of EC53 in both quiescence and outburst. The models include the contribution of external heating by the ambient radiation field and the different components of an embedded YSO: the central protostar, a circumstellar disc, envelope and bipolar cavities. Baek et al.~find that the SED of the system from quiescence to outburst is best modelled by an increase in luminosity from 6 $L_\odot$ to 20 $L_\odot$ for a system with outer envelope size, $R_{\rm env} = 10000$ au, radial density power-law index, $p$ = 1.5, and cavity opening angle, $\theta_{\rm cav} = 20$. Using this best fit 2-D model for the system parameters shown above and without consideration of external heating, we explore the relationship (or the value of $\eta$) between the WISE W2 and 850 $\mu$m emission by increasing the outburst luminosity by a factor 3.3 to 1000 (see Fig. 9). For this model, $\eta$ is found to be close to 4 independent of the outburst luminosity. Looking more closely we find that F$_{850} \propto$ L$^{0.28}$ which is very similar to the expected value derived in the preceding subsection, whereas the modelled mid-IR flux varies less strongly with luminosity than required to fit the observed sub-mm to mid-IR scaling.

In the above tests, the only source heating the disc is the central protostar (passive heating). If, however, the accretion rate is high enough, then the discs are also heated by viscous accretion \citep[as in FUor discs, e.g.][]{1996Hartmann}. Dust continuum modelling of the eruptive YSO V883 Ori \citep{2019Lee} shows that, within 10 au, the dust temperature may be higher at the disc midplane than the disc surface. To account for this effect in the Baek et al.\ models, the temperature of the disc midplane ($n_{\rm H_2}  > 10^{10}$ cm$^{-3}$) was raised within an arbitrary radius to three times that of the temperature estimated from passive heating alone. For radii where the increased temperature is greater than 1200 K (the dust dissipation temperature), then a temperature of 1200 K is adopted. This new effect is tested for four different luminosity increases, to 20, 60 ,120 and 180 $L_\odot$, where the boundary for viscous heating is set at 1, 2.5, 4 and 6 au, respectively. Additionally, for the models with luminosity increases to 20 and 120 $L_\odot$ we also tested boundary radii of 3 and 10 au respectively. Fig.\ 10 shows that $\eta$ for the six tested cases lies above the $\eta$ = 4 line.  The change is driven primarily by increasing the mid-IR emission, which WISE observes, in comparison to the far-IR, which remains responsible for heating the outer envelope (Fig.\ 10). Thus, the fractional importance of the viscously heated disc is stronger when the boundary for viscous heating is set to larger radii.

\subsection{Commentary}

The above arguments stress the fundamental importance to the resultant SED of a variety of physical parameters in the protostellar surroundings, including: the dust opacity law, at frequencies where dust emits and absorbs, the density structure within the envelope, especially near the location of the effective photosphere, and the shape of the radiation field. Additionally, adding a second luminous source, such as a viscously heated disc, affects the SED. Further investigation is required to understand how these parameters constrain the range of $\eta$. As importantly, the small range of observed $\eta$ values uncovered by this study suggest that these investigations will also provide useful constraints on the range of physical the parameters of the dust envelopes around protostars.

\begin{figure}
	\resizebox{\columnwidth}{!}{\includegraphics{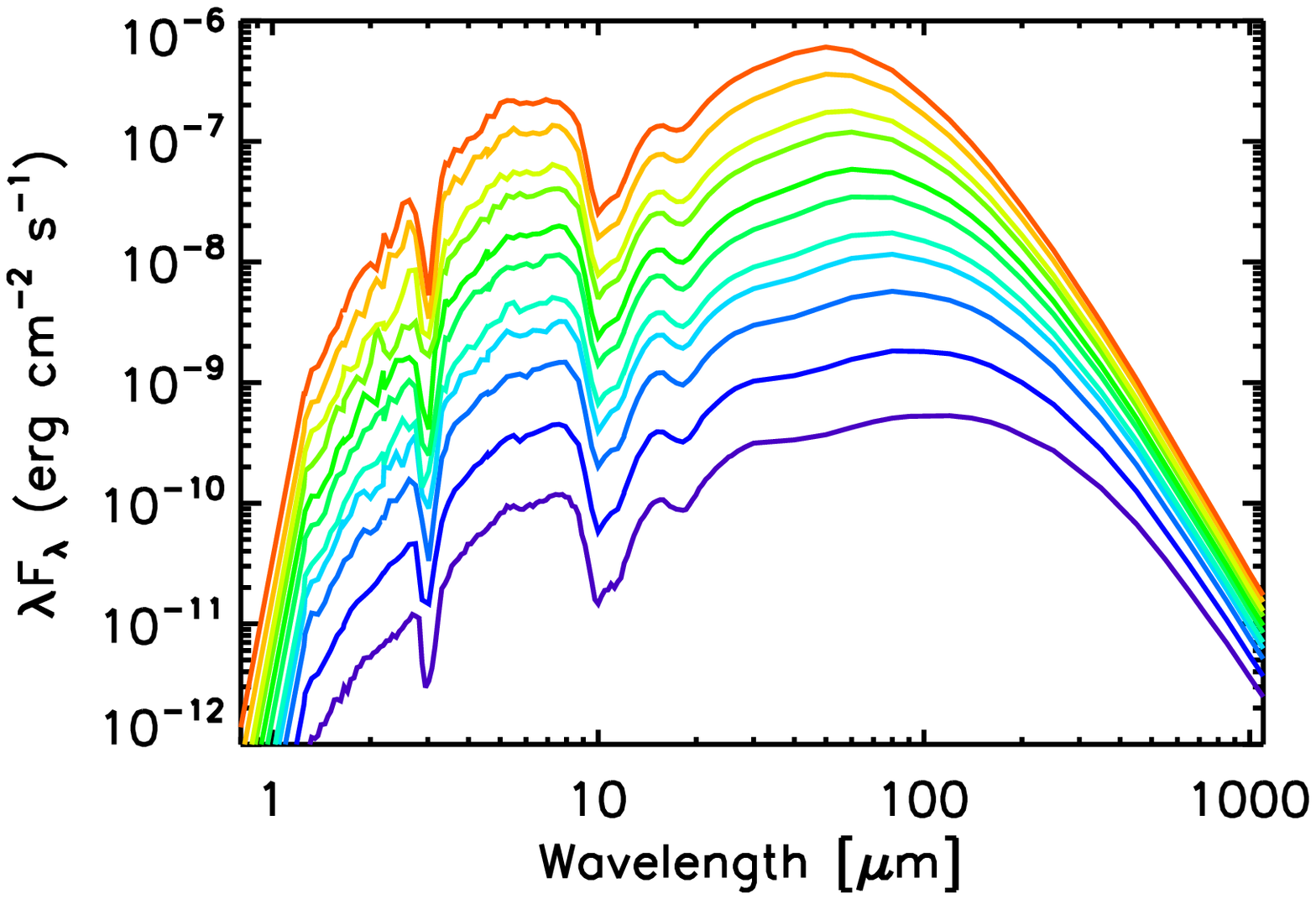}}\\
	\resizebox{\columnwidth}{!}{\includegraphics{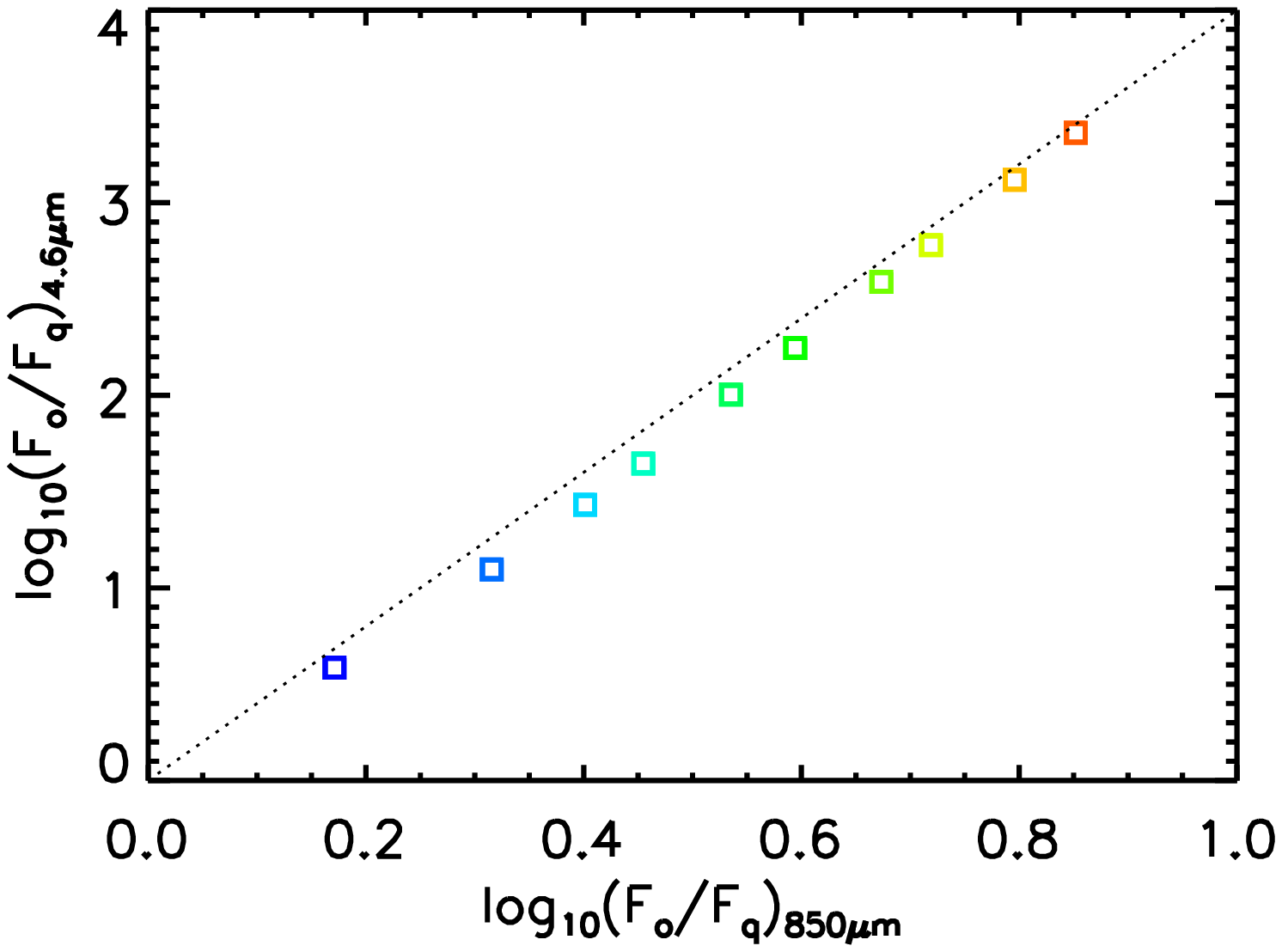}}
    \caption{(top)  SED models of EC53 that result from increasing the protostellar luminosity by a factor 3.3--1000 from the fiducial model of the system. (bottom) $\log$ (F/F$_{0}$) ({\it WISE}) vs $\log$ (F/F$_{0}$) (JCMT) for the various models of increase in protostellar luminosity, and for a disc inclination of 30$^{\circ}$. In the plot, the  dashed black line represents $\eta$=4.}
    \label{fig:baek1}
\end{figure}

\begin{figure}
	\resizebox{\columnwidth}{!}{\includegraphics{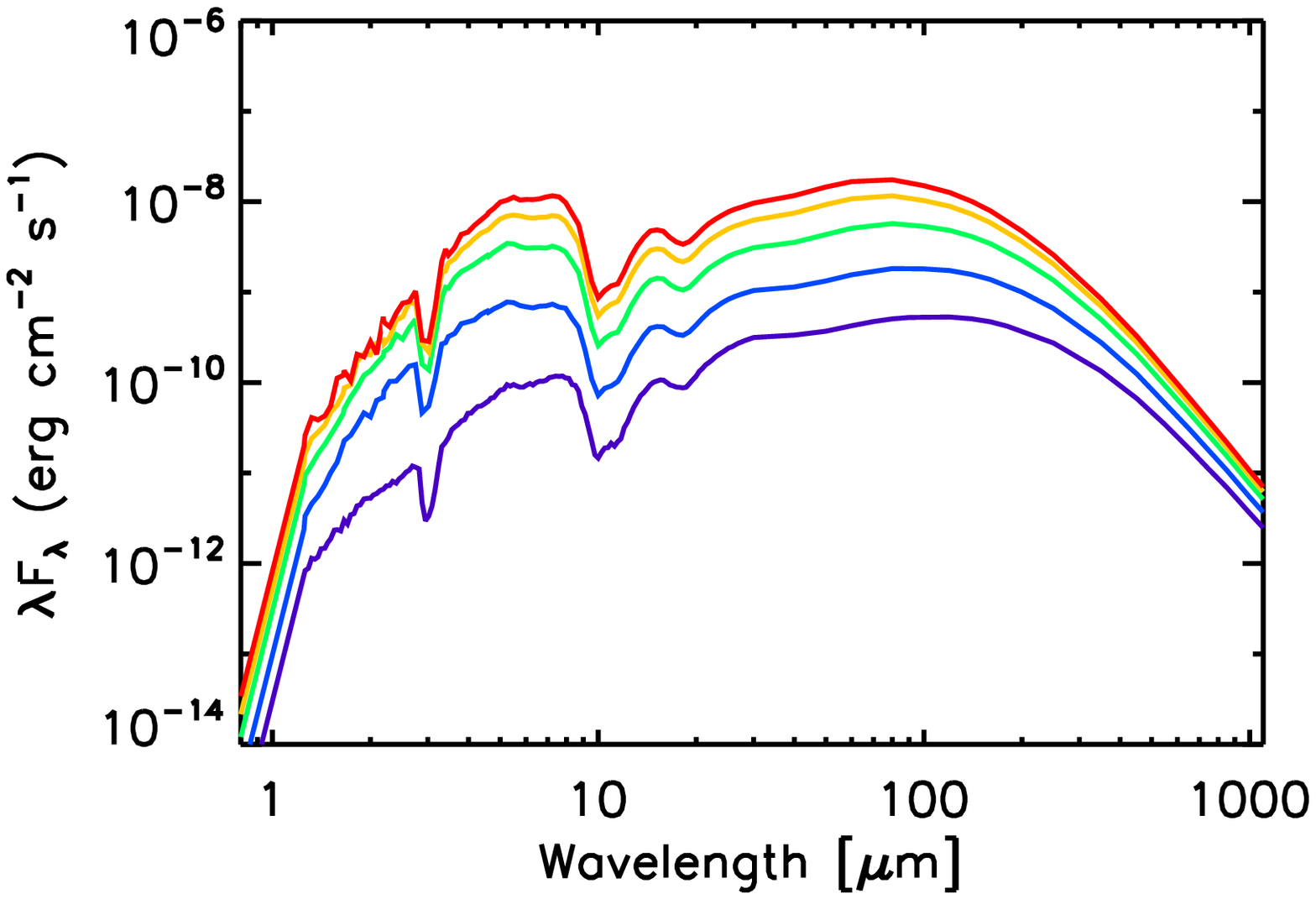}}\\
	\resizebox{\columnwidth}{!}{\includegraphics{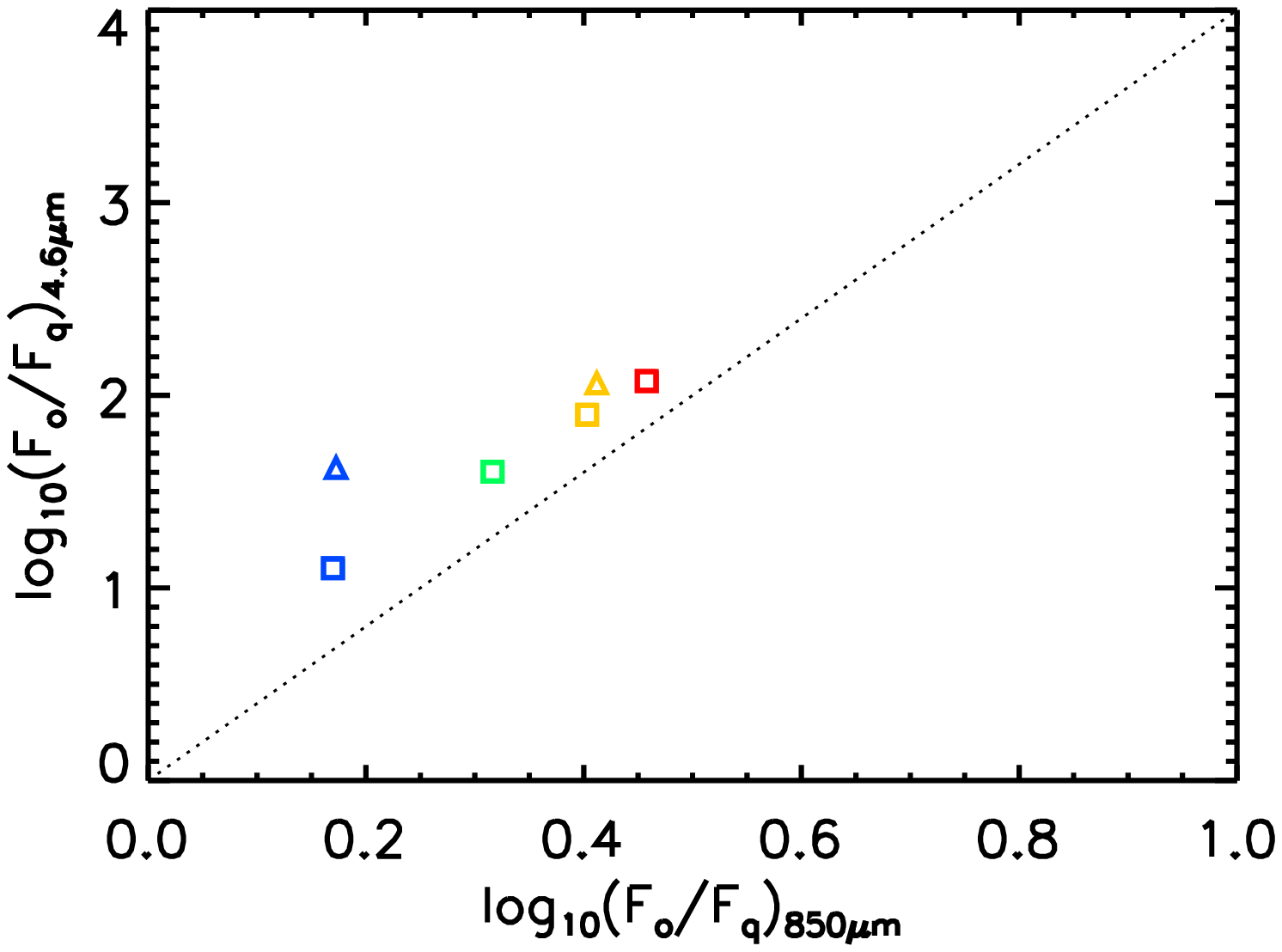}}
    \caption{ (top) SEDs for the fiducial model of EC53 at 6 L$_{\odot}$ (purple) and for increasing the luminosity to 20 (blue), 60 (green), 120 (orange) and 180 L$_{\odot}$ (red). (bottom) $\log$ (F/F$_{0}$) ({\it WISE}) vs $\log$ (F/F$_{0}$) (JCMT) for the different values of the luminosity. For each model we tested using different radii for the boundary of viscous heating. For the 20 L$_{\odot}$ radii of 1 au (blue square) and 3 au (blue triangle) are tested. For the 60 L$_{\odot}$ we set the boundary at 2.5 au (green square). Radii of 4 (orange square) and 10 au (orange triangle) are tested in the 120 L$_{\odot}$ model. Finally, a radius of 6 au is used for the 180 L$_{\odot}$ model. }
    \label{fig:baek2}
\end{figure}

\section{Summary}

We have studied the relationship between the mid-IR and sub-mm variability of deeply embedded protostars using the multi-epoch, contemporaneous data from the Wide Infrared Survey Explorer ({\it WISE}/NEOWISE) and the ongoing JCMT Transient Survey for a sample of 59 bright sub-mm sources with good {\it WISE data}.  We analysed the data from both surveys in the search for signs of stochastic (random) and/or secular (roughly monotonic in time) variability, recognizing that large changes in the accretion rate should lead to an increase of the observed flux across the spectrum of the YSO \citep{2013Scholz, 2018Johnstone,2019Macfarlane_a,2020Baek}.

We find that 16 out of 59 sources display variability at 850 $\mu$m. The majority of the YSO sample (33 out of 59 objects) is variable in {\it WISE} with a median amplitude of $\Delta W2=0.35$~mag, larger than the median amplitudes found for YSOs in star forming regions observed by the YSOVAR team. Given the expectation that the amplitude of variability increases toward early evolutionary stages \citep[e.g.][]{2017Contreras_a, 2014Gunther, 2018Wolk}, this is not surprising. The YSOs in our sample are associated with bright sub-mm sources, therefore we are selecting YSOs at earlier evolutionary stages compared to e.g. those observed by the YSOVAR team.

Since we were interested in studying the long-term variability arising from changes in the accretion rates of protostellar sources, our analysis focused only on the secular changes observed in the data from the JCMT and {\it WISE} surveys. In twenty-four objects we do not observe signatures of variability in either survey. In two cases we find that the sub-mm variability is not observed at mid-IR wavelengths. In one of these cases, this is likely explained by the low number of reliable data points, while for the second YSO it is hard to explain this behaviour; the mid-IR and sub-mm emission may arise from different sources.

For 19 YSOs, variability is observed at $4.6 \mu$m but not at 850 $\mu$m. For objects with a mean brightness at 850 $\mu$m lower than 0.4 Jy beam$^{-1}$, the sub-mm variability would be detected only for strong variables, with a change in {\it WISE} photometry of $\sim$2.5 magnitudes. In addition, low amplitude variability at mid-IR ($<$0.38 mag) will not be observed at 850$\mu$m independent of the brightness of the object. Four YSOs are above these brightness and amplitude limits but do not show sub-mm variability. In two of these sources, the mid-IR variability is either short-term or explained by variable extinction, and thus it does not affect emission at 850$\mu$m. In tother two YSOs, it appears more likely that the mid-IR and sub-mm emission does not arise from the same source.

For 14 YSOs we observe variability with linear trends at both mid-IR and sub-mm. We performed fits to the fluxes of the 14 YSOs showing secular changes at both mid-IR and sub-mm.  The time dependence of the fluxes at mid-IR and sub-mm wavelengths follow a relation of $\log_{10} F_{4.6}(t)=\eta \log_{10} F_{850}(t)$, with $\eta = 5.53 \pm 0.29$. 
Using the SED models of Baek et al., we find that the value of $\eta$ remains close to four for the envelope parameters that best fit the source EC53, suggesting that something is still missing from the envelope modelling.  We suggest that the larger observed value of $\eta$ may be explained by a combination of dust opacity and envelope density structure, along with the possible addition of a second heating source, such as a viscously heated disc.

The results of this work show that contemporaneous observations across the spectrum of deeply embedded YSOs gives us a probe into the interior structure of these systems, regions that are otherwise difficult to understand.  Future sub-mm observations in conjunction with infrared missions from the James Webb Space Telescope \citep[JWST;][]{2006Gardner} and the SPace Infrared telescope for Cosmology and Astrophysics \citep[SPICA;][]{2018Roelfsema,2019Andre} will provide further insights into the structure of deeply embedded YSOs.

\section*{Acknowledgements}

The authors thank the anonymous referee for the useful comments that helped to improve the manuscript.

This publication makes use of data products from the Near-Earth Object Wide-field Infrared Survey Explorer (NEOWISE), which is a project of the Jet Propulsion Laboratory/California Institute of Technology. NEOWISE is funded by the National Aeronautics and Space Administration. This research has made use of the NASA/ IPAC Infrared Science Archive, which is operated by the Jet Propulsion Laboratory, California Institute of Technology, under contract with the National Aeronautics and Space Administration. The contribution of C.C.P. was funded by a Leverhulme Trust Research Project Grant. A.S. is supported by the STFC grant no. ST/R000824/1. G.J.H. is supported by general grant 11773002 awarded by the National Science Foundation of China. D.J. is supported by NRC Canada and by an NSERC Discovery Grant. J.-E. L. and G. B. are supported by the Basic Science Research Program through the National Research Foundation of Korea (grant No. NRF-2018R1A2B6003423) and the Korea Astronomy and Space Science Institute under the R\&D program supervised by the Ministry of Science, ICT and Future Planning. G.B. was also supported by the National Research Foundation of Korea (NRF) Grant funded by the Korean Government (NRF-2017H1A2A1043046-Global Ph.D. Fellowship Program).

The authors thank the JCMT staff for their support of the GBS team in data collection and reduction efforts.

The James Clerk Maxwell Telescope has historically been operated by the Joint Astronomy Centre on behalf of the Science and Technology Facilities Council of the United Kingdom, the National Research Council of Canada and the Netherlands Organisation for Scientific Research

The authors wish to recognise and acknowledge the very significant cultural role and reverence that the summit of Maunakea has always had within the indigenous Hawaiian community. We are most fortunate to have the opportunity to conduct observations from this mountain. The James Clerk Maxwell Telescope is operated by the East Asian Observatory on behalf of The National Astronomical Observatory of Japan; Academia Sinica Institute of Astronomy and Astrophysics; the Korea Astronomy and Space Science Institute; the Operation, Maintenance and Upgrading Fund for Astronomical Telescopes and Facility Instruments, budgeted from the Ministry of Finance (MOF) of China and administrated by the Chinese Academy of Sciences (CAS), as well as the National Key R\&D Program of China (No. 2017YFA0402700). Additional funding support is provided by the Science and Technology Facilities Council of the United Kingdom and participating universities in the United Kingdom and Canada. Additional funds for the construction of SCUBA-2 were provided by the Canada Foundation for Innovation. This research used the facilities of the Canadian Astronomy Data Centre operated by the National Research Council of Canada with the support of the Canadian Space Agency. This research has made use of the SIMBAD database, operated at CDS, Strasbourg, France (Wenger et al. 2000).





\bibliographystyle{mnras}




\appendix

\section{{\it WISE} saturated sources}\label{app:sat}

In Section \ref{sec:method} we established that 24 out of 307 JCMT bright sources have {\it WISE} detections (within 10 arcsec) that are saturated range in both $W1$ and $W2$. We determine that 21 of these sources are associated with known protostars. Given this, we analysed the saturated objects to check whether we are missing any variable sources that could be useful additions in the determination of the relationship between the mid-IR and the sub-mm variability of YSOs.

The visual inspection of the 24 sources reveals that in 16 objects the {\it WISE} and JCMT fluxes likely arise from the same source. We search for statistical signatures of variability in these 16 objects following the same method of Section \ref{sec:statvar}, with fluxes corrected for saturation following the guidance from the {\it WISE} supplementary material \citep{2012Cutri}.

\begin{figure*}
	\resizebox{1.5\columnwidth}{!}{\includegraphics{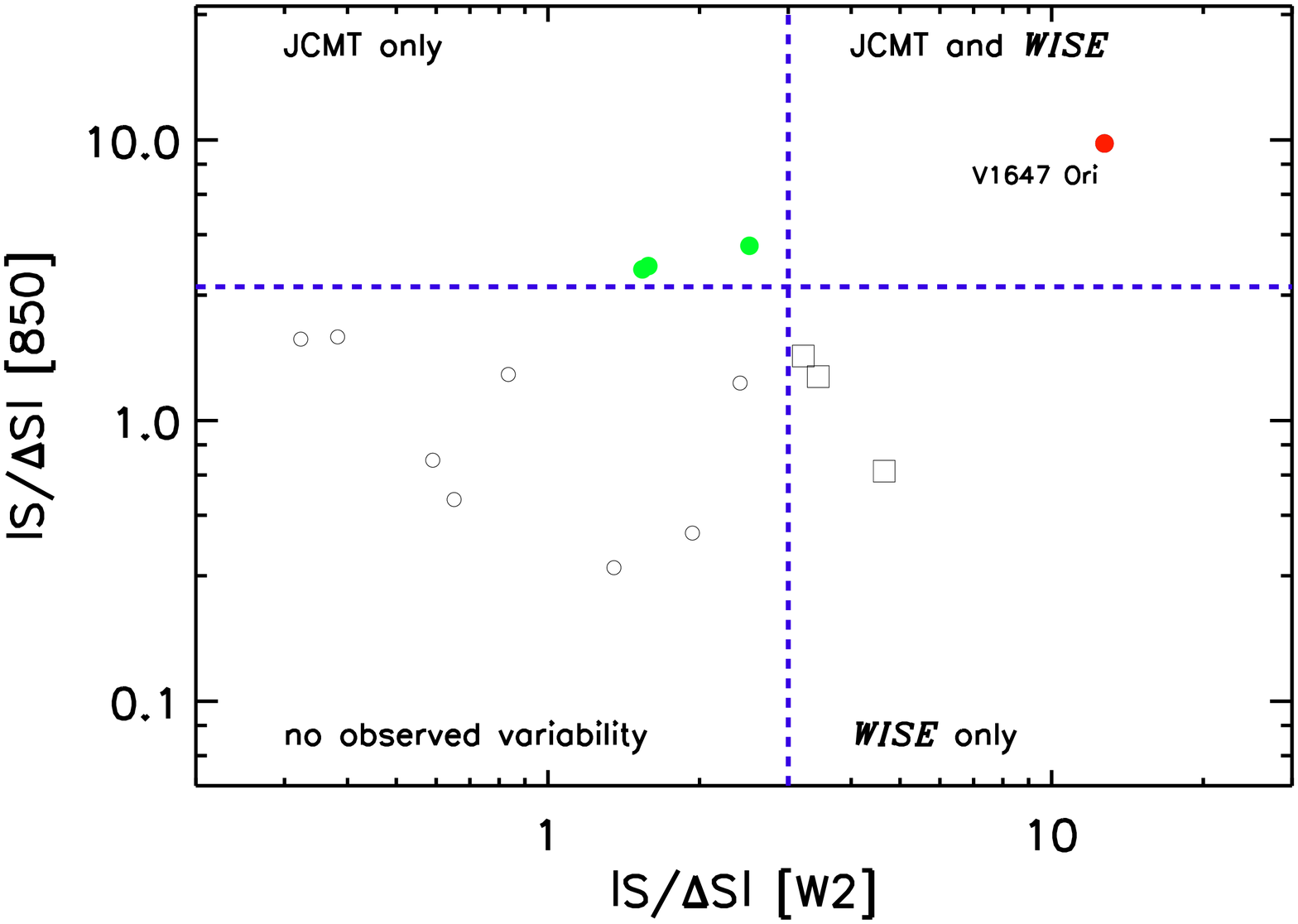}}\\
	\vspace{1cm}
	\resizebox{1.5\columnwidth}{!}{\includegraphics{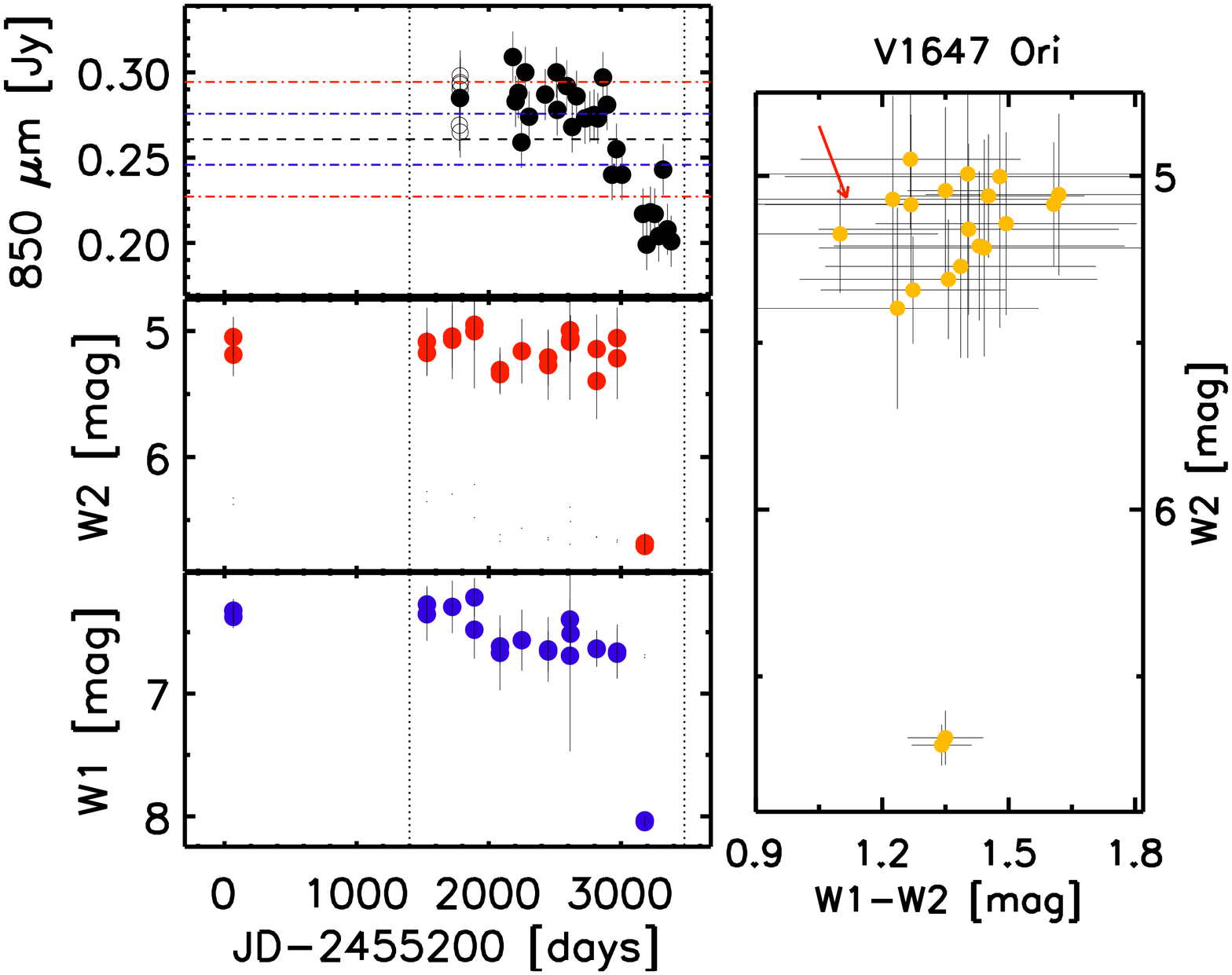}}
    \caption{(top) $|S/\Delta S|$ (JCMT) vs $|S/\Delta S|$ ({\it WISE} $W2$) for the 16 targets in our sample that are saturated in {\it WISE}. The blue dotted lines mark $|S/\Delta S|=3$. Non variable objects are marked with open circles, the upper left quadrant shows objects with variability only in the JCMT Transient survey (green circles), while the bottom right quadrant defines the region where we only observe variability at $W2$ (open squares). Finally objects that are found to be variable in both surveys are located in the upper right quadrant (solid circles). Objects with the largest variability in both surveys $|S/\Delta S| \geq6$  are marked by the solid red circles. (bottom) {\it WISE} $W1$, $W2$ magnitudes and JCMT 850 $\mu$m flux, and $W2$ vs $W1-W2$ plot for source 32 in NGC2068 (V1647 Ori). Symbols are the same as in Fig. \ref{fig:n1333_17}}
    \label{fig:v1647ori}
\end{figure*}

Fig. \ref{fig:v1647ori} shows the {\it WISE} $W2$ vs JCMT  statistical measure of secular variability ($|S/\Delta S|$) for the 16 objects. We find 4 of these YSOs show secular variability in the JCMT data ($|S/\Delta S| \geq3$):  source 2 in NGC1333 ([SVS76] NGC1333 13A \citealt{1976Strom}), source 16 in Ophiucus \citep[Elia 2--33][]{1978Elias}, source 12 in Serpens Main \citep[IRAS 18274$+$0112][]{1976Strom2} and source 32 in NGC2068 \citep[V1647 Ori][]{2004Mcneil}. In the first three cases, this variability is not detected in {\it WISE}, thus they are found in the JCMT only region of the figure. The high saturation of the {\it WISE} emission of objects IRAS 18274$+$0112 and Elia 2--33 lead to large errors in the photometry, making it impossible to determine any mid-IR variability. YSO [SVS76] NGC133313A shows a long-term linear increase at 850 $\mu$m, while the mid-IR lightcurve shows an apparent fading. The most likely explanation is that the JCMT and {\it WISE} fluxes are not arising from the same source.

In only one case, V1647 Ori, we find that variability in both JCMT and {\it WISE} data. This Class I YSO \citep{2012Megeath} is the illuminating star of the McNeil's nebula and  is a known eruptive variable  \citep[see e.g.][]{2018Connelley}. The source shows a large decrease in brightness at both 850 $\mu$m and the mid-IR (see Fig. \ref{fig:v1647ori}). The inspection of single exposure $W2$ images from the NEOWISE survey confirms that the source has faded by 1.8 mag at 4.6 $\mu$m. Given the observed variability, we include this source in the analysis of Section \ref{sec:wjvar}.

\section{Individual notes}\label{sec:ind}

\subsection{Stochastic}\label{sec:ind_sto}

\begin{figure*}
	\resizebox{1.7\columnwidth}{!}{\includegraphics{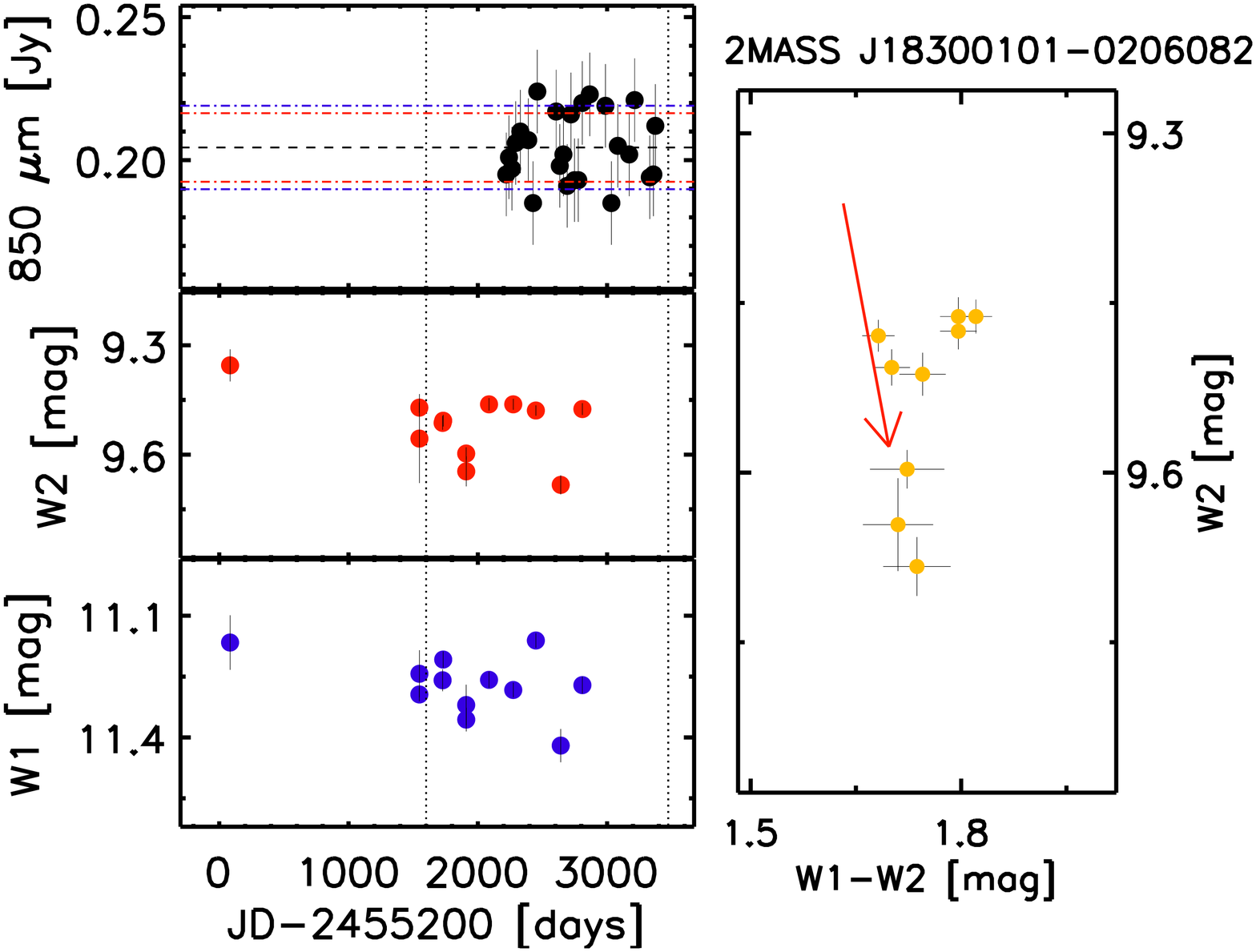}}
    \caption{{\it WISE} $W1$, $W2$ magnitudes and JCMT 850 $\mu$m flux, and $W2$ vs $W1-W2$ plot for source 13 in Serpens South. Symbols are the same as in Fig. \ref{fig:n1333_17}}
    \label{fig:oc56}
\end{figure*}

Source 13 in Serpens South \citep[2MASS J18300101$-$0206082;][]{2015Dunham} shows stochastic variability but with no long-term trends in the {\it WISE} $W2$ data (Fig. \ref{fig:oc56}). This object does not show any signs of variability in the JCMT Transient survey. In Fig. \ref{fig:oc56} we can see that the variability is low amplitude ($\Delta W2=0.2$~mag)  and that $W1-W2$ colour change is consistent with variable extinction, a mechanism that should not affect the $850 \mu$m flux. This object is also faint at 850 $\mu$m, so any corresponding variability would be below our detection limits.

\subsection{Secular}

\subsubsection{MHO 3271}\label{sec:ss11}

\begin{figure*}
	\resizebox{1.7\columnwidth}{!}{\includegraphics{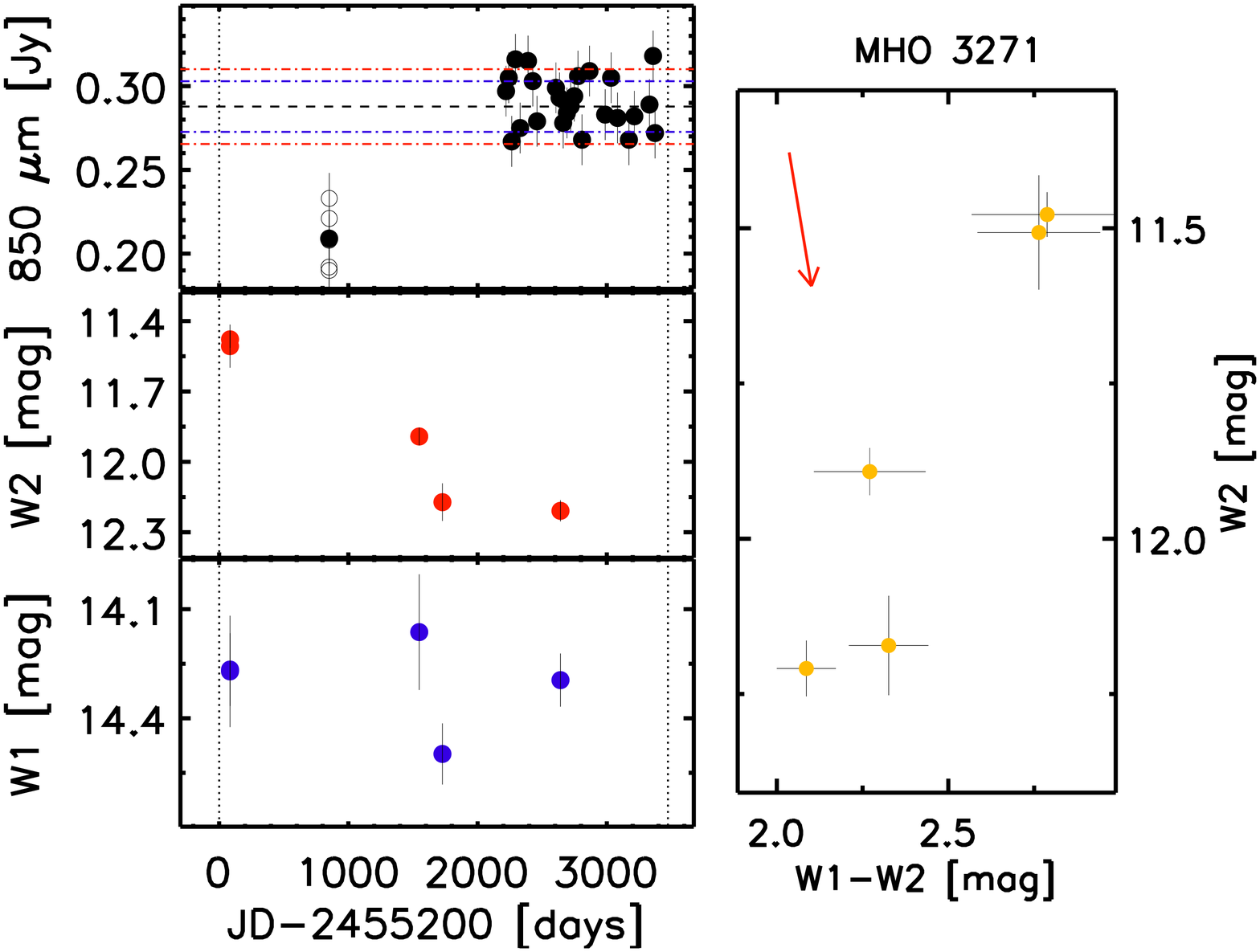}}\\
    \caption{{\it WISE} $W1$, $W2$ magnitudes, 850 $\mu$m flux from the JCMT, and $W2-W2$ vs $W1$ for source 11 in Serpens South. Symbols are the same as in Fig. \ref{fig:n1333_17}}
    \label{fig:ss11}
\end{figure*}

Source 11 in Serpens South \citep[MHO 3271;][]{2015Zhang} is a Class I YSO that shows secular variability in both surveys. The measured S/$\Delta$S from both surveys indicates that the variability is anti-correlated. While the object brightens between the GBS and Transient survey observations, the $W2$ lightcurve appears to fade during this time (see Fig \ref{fig:ss11}). However, inspection of the light curves show this conclusion is based only on a handful of epochs. In addition, only one epoch in $W2$ is contemporaneous to the JCMT data. With such little overlap, we did not include this object in any further analysis in the main section of the paper.

\subsubsection{JCMT only}\label{sec:ind_jcmt}

In the top left corner of Fig. \ref{fig:ex3}, two objects, source 5 in NGC2068 \citep[HBC 502][]{1988Herbig} and source 0 in NGC1333 \citep[IRAS4A][]{1987Jennings}, show significant secular variability in JCMT but not in {\it WISE}. 

\begin{figure*}
	\resizebox{1.7\columnwidth}{!}{\includegraphics{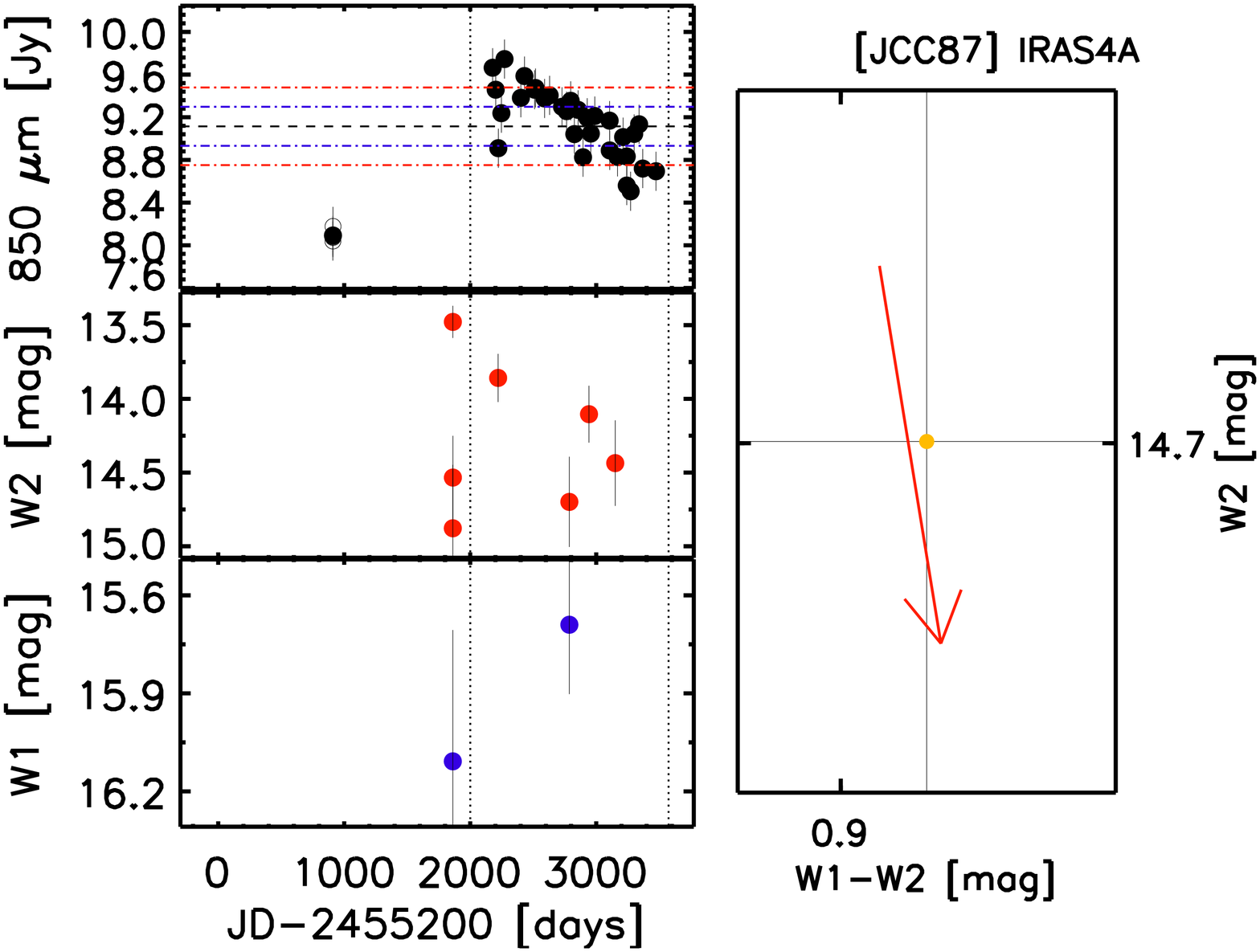}}\\
	\vspace{0.5cm}
	\resizebox{1.7\columnwidth}{!}{\includegraphics{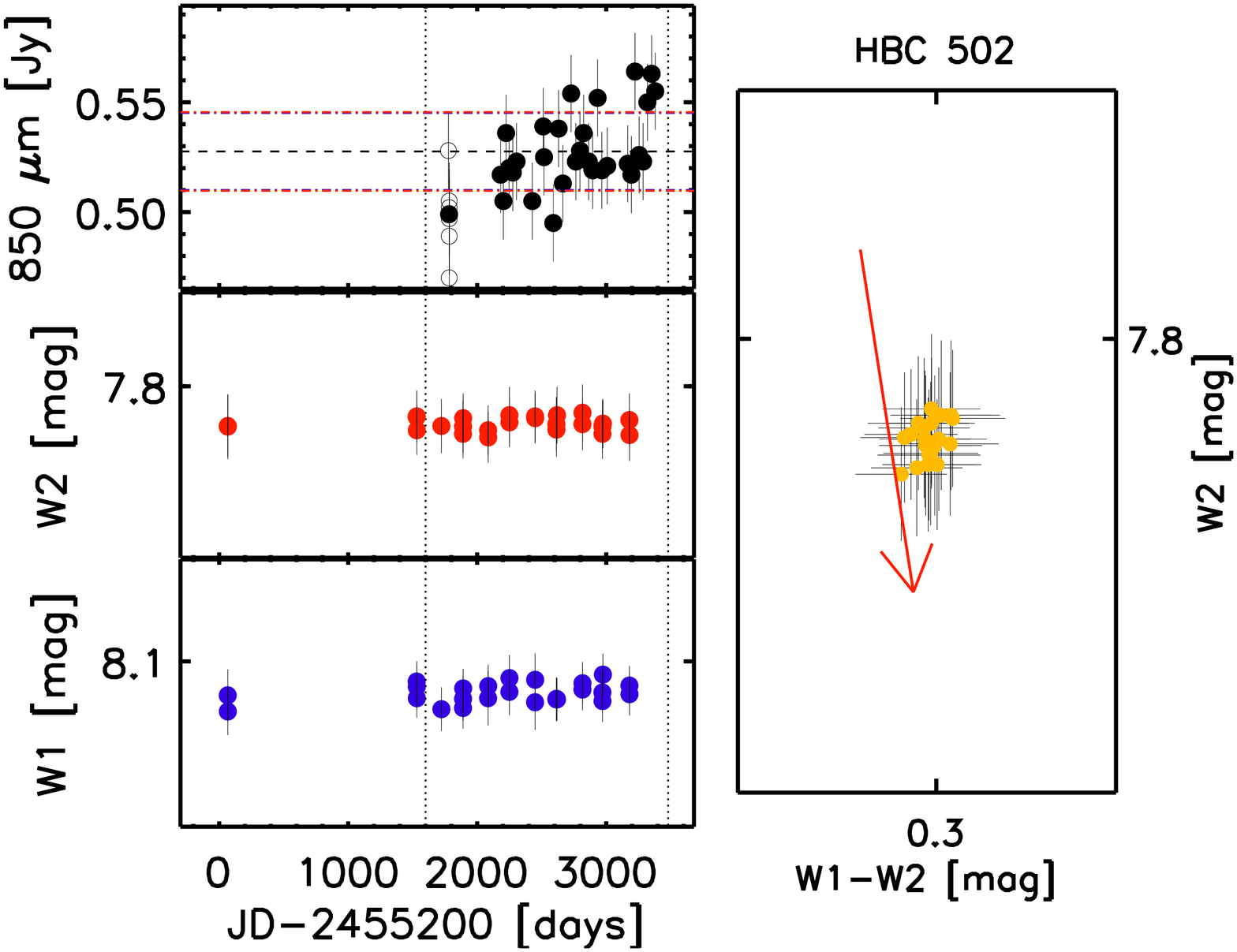}}
    \caption{{\it WISE} $W1$, $W2$ magnitudes, 850 $\mu$m flux from the JCMT, and $W2-W2$ vs $W1$ for source 0 in NGC1333 (top) and for source 5 in NGC2068 (bottom). Symbols are the same as in Fig. \ref{fig:n1333_17}}
    \label{fig:jcmt_only}
\end{figure*}

For IRAS4A, the lack of {\it WISE} $W2$ variability may be caused by the low number of reliable data points (see Fig. \ref{fig:jcmt_only}). However, the {\it WISE} photometry (Fig. \ref{fig:jcmt_only}) confirms that this object brightened between the GBS and Transient survey observations \citep[see also][]{2017Mairs}, as the source is not detected in the original {\it WISE} mission, and only goes above the detection limit of {\it NEOWISE} when the object is at its maximum brightness.

HBC 502 shows long-term linear variability in the GBS and transient survey data. Surprisingly, the $W2$ and $W1$ light curves of the source show that the object remains at an approximately constant magnitude between the {\it WISE} and {\it NEOWISE} surveys (see Fig. \ref{fig:jcmt_only}). The coordinates of the mid-IR detections are within 1 arcsec of the coordinates of class II YSO HBC 502, and 6 arcsec from the coordinates of the 850 $\mu$m peak. The visual inspection of Section \ref{sec:visual} does not show evidence that the {\it WISE} and JCMT detections are unrelated. However, it is hard to explain the observation of variability at 850 $\mu$m but not in the $W1$ and $W2$ bands, especially since the source is bright in both surveys. The most likely explanation is that the JCMT peak is not associated with HBC 502.

\subsubsection{{\it WISE} only}\label{sec:ind_wise}

Four objects that are variable in {\it WISE} but not in JCMT data are located above the brightness and amplitude limits discussed in Section \ref{sec:wisevar}. These objects correspond to source 18 in NGC1333 \citep[Class I YSO SSTc2d J032901.6$+$312021][]{2013Dunham}, source 6 in OMC2/3 \citep[Class 0 YSO HOPS 60][]{2016Furlan}, source 4 in NGC2068 \citep[Class I YSO HOPS315][]{2016Furlan} and source 35 in NGC1333 \citep[Class I YSO SSTc2dJ032837.1$+$311331][]{2013Dunham}. 

\begin{figure*}
	\resizebox{1.7\columnwidth}{!}{\includegraphics{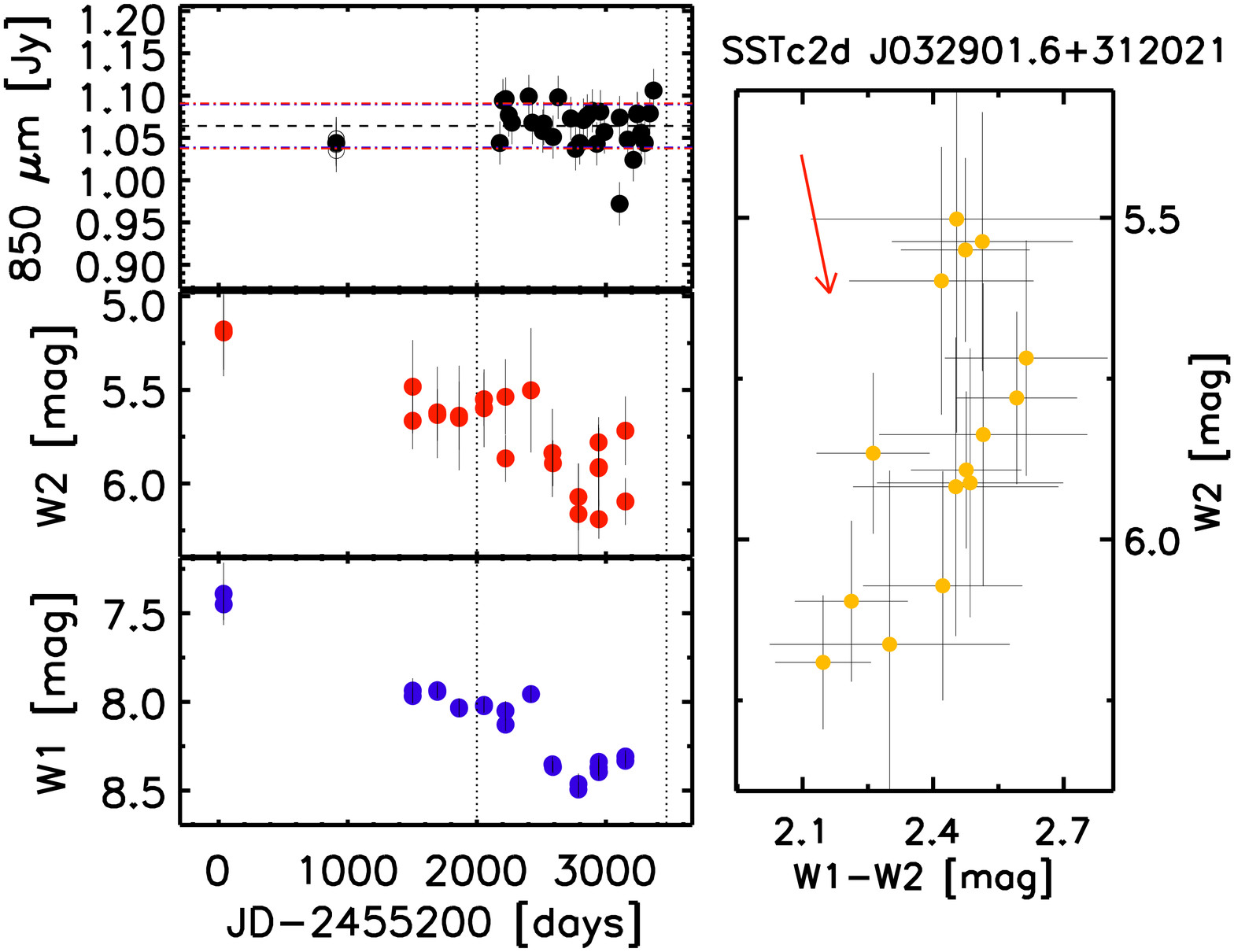}}\\
	\vspace{0.5cm}
	\resizebox{1.7\columnwidth}{!}{\includegraphics{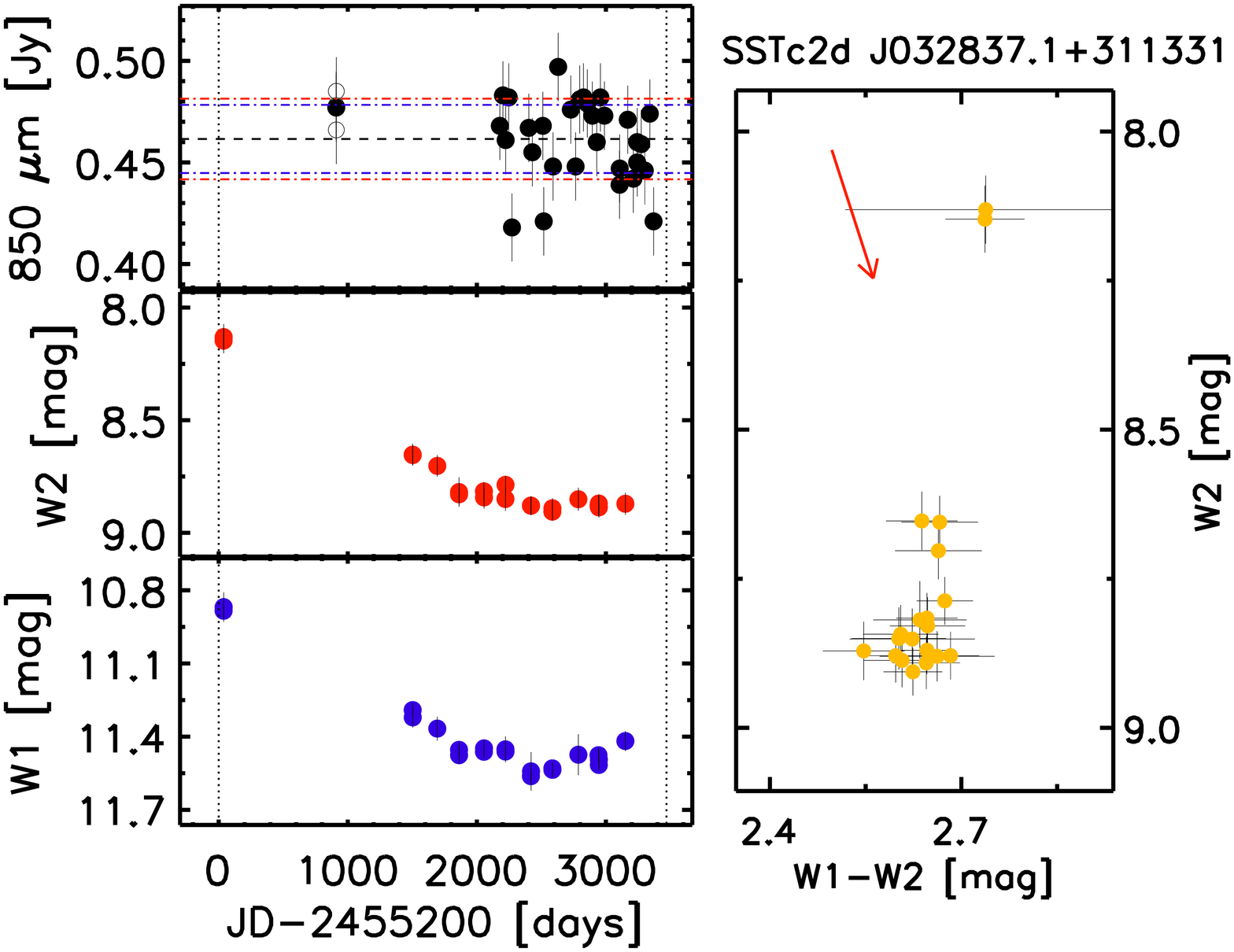}}
\caption{{\it WISE} $W1$, $W2$ magnitudes, 850 $\mu$m flux from the JCMT, and $W2-W2$ vs $W1$ for source 18 in NGC1333 (top) and for source 35 in NGC1333 (bottom). Symbols are the same as in Fig. \ref{fig:n1333_17}.}
    \label{fig:4sourcesa}
\end{figure*}

\begin{figure*}
	\resizebox{1.7\columnwidth}{!}{\includegraphics{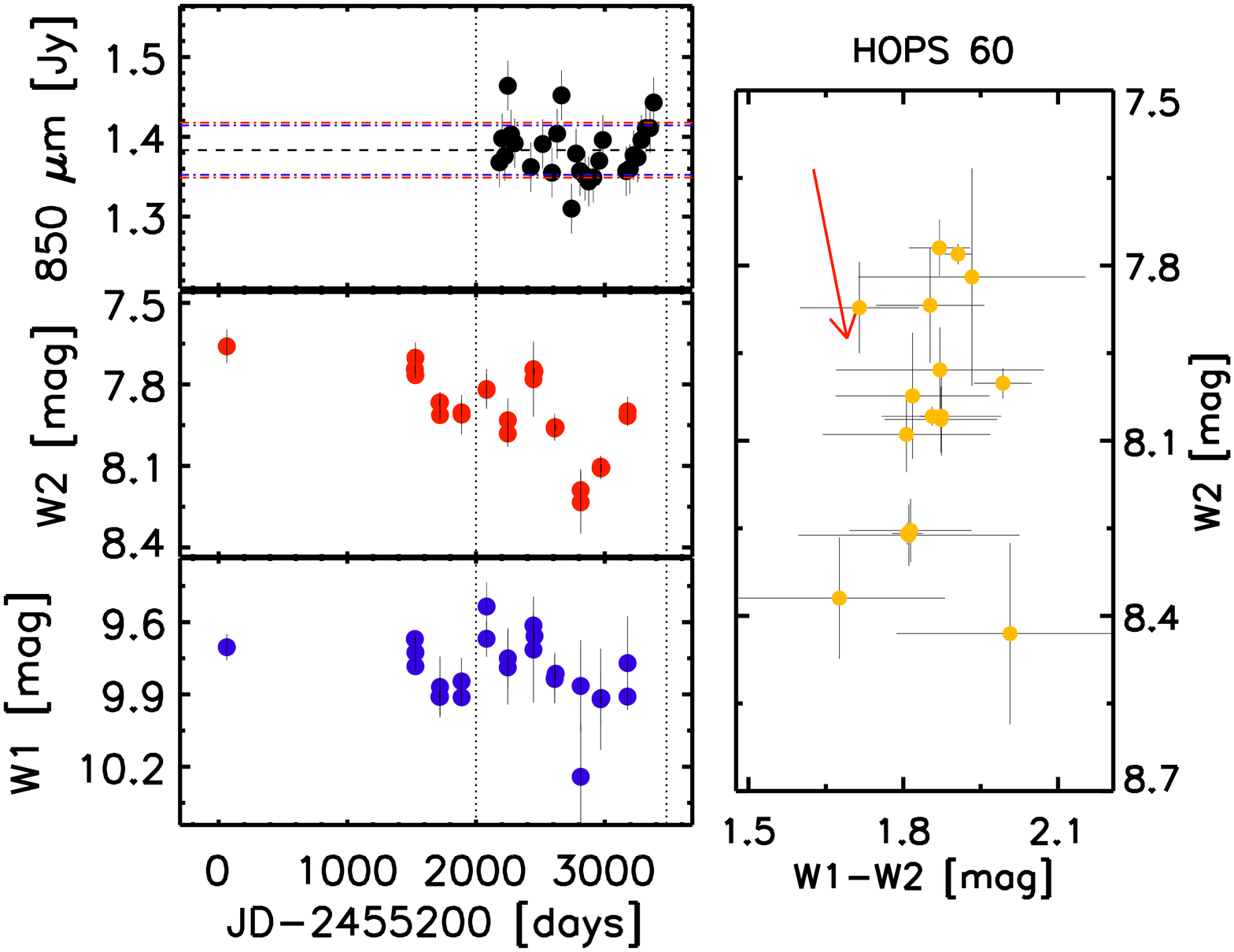}}\\
	\vspace{0.5cm}
	\resizebox{1.7\columnwidth}{!}{\includegraphics{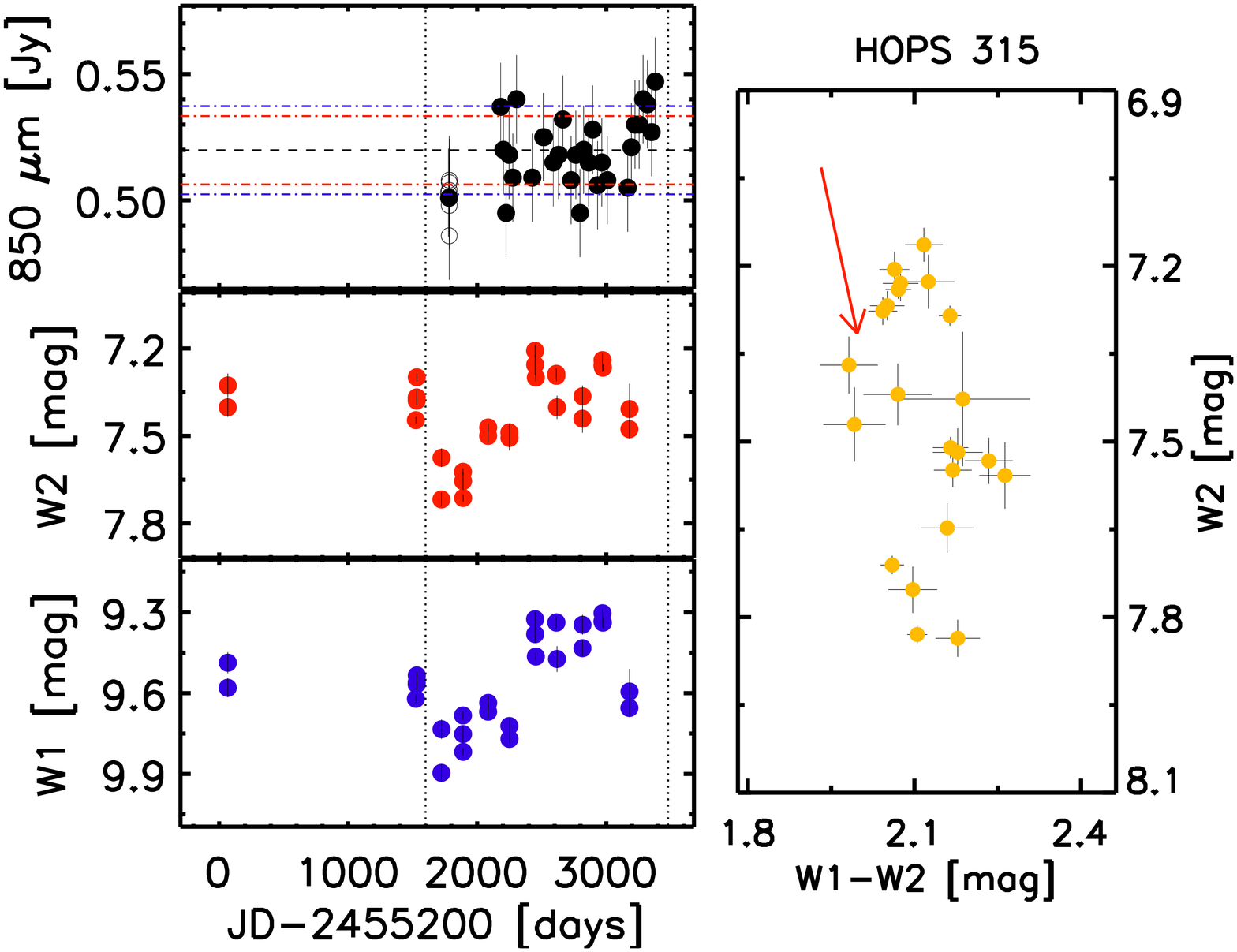}}
   \caption{{\it WISE} $W1$, $W2$ magnitudes, 850 $\mu$m flux from the JCMT, and $W2-W2$ vs $W1$ for source 6 in OMC2/3 (top), source 4 in NGC2068 (bottom). Symbols are the same as in Fig. \ref{fig:n1333_17}.}
    \label{fig:4sourcesb}
\end{figure*}

Figs. \ref{fig:4sourcesa} and \ref{fig:4sourcesb} show the {\it WISE} and JCMT light curves, as well as the $W2$ vs $W1-W2$ change, for the five objects. For sources HOPS 315 and HOPS 60, the {\it WISE} variability seems to be driven by short-term events that do not appear to have an effect on the sub-mm emission of the system. The variability of HOPS 315 seems to follow the reddening vector, thus we would not expect such variability to affect the sub-mm emission of the source.

Sources SSTc2d J032901.6$+$312021 and SSTc2dJ032837.1$+$311331 in NGC1333 show long-term declines that are not consistent with an increase in extinction along the line of sight, as they do not follow the reddening vector in $W2$ vs $W1-W2$ plots. However, this variability does not correlate with the sub-mm emission. Similar to the case of HBC 502 in Appendix \ref{sec:ind_jcmt} it is hard to explain this behaviour. The most likely explanation seems to be that the {\it WISE} and JCMT sources are not related.

\bsp	
\label{lastpage}
\end{document}